\documentclass[11pt,a4paper]{article}
\pdfoutput=1

\usepackage{jheppub}
\usepackage{latexsym}
\usepackage{revsymb}
\usepackage{multirow}
\usepackage{color}
\usepackage[usenames,dvipsnames,svgnames,table]{xcolor}

\usepackage{graphicx}
\usepackage{epsfig}  
\usepackage{epsf}    
\usepackage{dcolumn}
\usepackage{bm}
\usepackage{dcolumn}
\usepackage{textcomp}
\usepackage{float}
\usepackage{hypcap}
\usepackage[]{hyperref}
\usepackage{adjustbox}
\usepackage{multirow}

\usepackage{amsmath}
\usepackage{subfigure}
\usepackage{alphalph}
\usepackage{morefloats}
\usepackage{textcomp}
\usepackage[utf8x]{inputenc}
\usepackage{float}
\restylefloat{table}

\newcommand{\be}{\begin{equation}}
\newcommand{\ee}{\end{equation}}
\newcommand{\ba}{\begin{eqnarray}}
\newcommand{\ea}{\end{eqnarray}}

\newcommand{\cnv}{Cherenkov}

\title{Exploring invisible neutrino decay at ESSnuSB} 

\author[a,b]{Sandhya Choubey,}
\author[a,b,c]{Monojit Ghosh,}
\author[a,b]{Daniel Kempe,}
\author[a,b,d]{Tommy Ohlsson}

\affiliation[a]{Department of Physics, School of Engineering Sciences, KTH Royal Institute of Technology,\\ AlbaNova University Center, Roslagstullsbacken 21, SE--106 91 Stockholm, Sweden }
\affiliation[b]{The Oskar Klein Centre, AlbaNova University Center, Roslagstullsbacken 21,\\ SE--106 91 Stockholm, Sweden}
\affiliation[c]{Center of Excellence for Advanced Materials and Sensing Devices, Ruder Bo\v{s}kovi\'c Institute, 10000 Zagreb, Croatia}
\affiliation[d]{University of Iceland, Science Institute, Dunhaga 3, IS--107 Reykjavik, Iceland}

\emailAdd{choubey@kth.se}
\emailAdd{manojit@kth.se}
\emailAdd{dkempe@kth.se}
\emailAdd{tohlsson@kth.se}

\abstract{
We explore invisible neutrino decay in which a heavy active neutrino state decays into a light sterile neutrino state and present a comparative analysis of two baseline options, 540~km and 360~km, for the ESSnuSB experimental setup. Our analysis shows that ESSnuSB can put a bound on the decay parameter $\tau_3/m_3 = 2.64~(1.68) \times 10^{-11}$~s/eV for the baseline option of 360~(540)~km at $3 \sigma$. The expected bound obtained for 360~km is slightly better than the corresponding one of DUNE for a charged current (CC) analysis. Furthermore, we show that the capability of ESSnuSB to discover decay, and to measure the decay parameter precisely, is better for the baseline option of 540~km than that of 360~km. Regarding effects of decay in $\delta_{\rm CP}$ measurements, we find that in general the CP violation discovery potential is better in the presence of decay. The change in CP precision is significant  if one assumes decay in data but no decay in theory.
}

\arxivnumber{}
\begin{document}
\maketitle

\section{Introduction}
\label{sec1}

The phenomenon of neutrino oscillations in the standard three-flavor scenario can be expressed by three mixing angles $\theta_{12}$, $\theta_{23}$, and $\theta_{13}$, two mass-squared differences $\Delta m^2_{21} = m_2^2 - m_1^2$ and $\Delta m^2_{31} = m_3^2 - m_1^2$, and one Dirac-type CP-violating phase $\delta_{\rm CP}$.  During the past decades, data from solar, atmospheric, accelerator, and reactor neutrino experiments have successfully been able to determine the values of the parameters $\theta_{12}$, $\theta_{13}$, $\Delta m^2_{21}$, and $|\Delta m^2_{31}|$ to an excellent precision. The parameters, which are unknown at this moment, are: (i)~the mass ordering of neutrinos, i.e., $\Delta m^2_{31} > 0$ known as normal ordering (NO) or $\Delta m^2_{31} < 0$ known as inverted ordering (IO), (ii)~the octant of the mixing angle $\theta_{23}$, i.e., $\theta_{23} > 45^\circ$ known as the higher octant (HO) or $\theta_{23} < 45^\circ$ known as the lower octant (LO), and (iii)~the true value of $\delta_{\rm CP}$ and its precision. At this moment, data from ongoing experiments provide a hint towards the true ordering as NO, the true octant as HO, and $\delta_{\rm CP} = -90^\circ$ \cite{Esteban:2020cvm}. There are many dedicated experiments to establish these hints concretely. 

Apart from standard neutrino oscillation physics, there are many new physics scenarios which can be probed at neutrino oscillation experiments. Invisible neutrino decay is an example of one such scenario \cite{Lindner:2001fx}. In invisible neutrino decay, a heavy neutrino state decays into a light neutrino state, which is sterile and therefore invisible\footnote{Neutrinos could also undergo visible decay, where the final neutrino state is an active neutrino.}. Theoretically, for Dirac neutrinos, this scenario can arise if there exists a coupling between the neutrinos and a light scalar boson \cite{Acker:1993sz}. In this case, decay can be defined as $\nu_j \rightarrow \nu_{i R} + \chi$, where $\nu_{i R}$ is a right-handed singlet and $\chi$ is an iso-singlet scalar. For Majorana neutrinos having a pseudo-scalar coupling with a Majoron $J$ and a sterile neutrino $\nu_s$, it is possible that $\nu_j \rightarrow \nu_s + J$ \cite{Chikashige:1980ui}. Irrespective of the model, in the presence of neutrino decay, the Hamiltonian for neutrino propagation is modified, and therefore, one can measure the decay parameter as well as the effect of decay on the measurement of the standard oscillation parameters in a neutrino oscillation experiment. In principle, all three neutrino states $\nu_1$, $\nu_2$, and $\nu_3$ can decay invisibly. The decay due to $\nu_2$ is severely constrained from solar neutrino data \cite{Bandyopadhyay:2002qg} and the constraints from supernova SN1987A \cite{Hirata:1987hu} is applicable to decays of $\nu_2$ and $\nu_1$ \cite{Frieman:1987as}. In Ref.~\cite{Lindner:2001th}, decays of supernova neutrinos have been investigated in great theoretical detail. Decay due to $\nu_3$ can be measured in present and future accelerator, atmospheric, and reactor neutrino experiments. Recently, a lot of work has been performed in this direction. Studies of invisible decay in the accelerator neutrino experiments T2K \cite{Abe:2020vdv}, NO$\nu$A \cite{Nosek:2019vls}, MINOS \cite{Michael:2008bc}, DUNE \cite{Abi:2020evt}, and MOMENT \cite{Cao:2014bea} can be found in Refs. \cite{Choubey:2018cfz,Gomes:2014yua,Choubey:2017dyu,Ghoshal:2020hyo,Tang:2018rer}. For the study of invisible neutrino decay in the ongoing atmospheric neutrino experiment Super-Kamiokande (SK) \cite{Hosaka:2006zd} and the future atmospheric neutrino experiment INO \cite{Kumar:2017sdq}, see Ref.~\cite{GonzalezGarcia:2008ru,Choubey:2017eyg,Mohan:2020tbi}, for the study with atmospheric neutrino data of the future ultra-high energy neutrino experiment KM3NeT-ORCA \cite{Adrian-Martinez:2016fdl}, see Ref.~\cite{deSalas:2018kri}, and for the medium baseline reactor neutrino experiment JUNO \cite{An:2015jdp}, see Ref.~\cite{Abrahao:2015rba}.

In this paper, we study the scenario of invisible decay of the neutrino state $\nu_3$ in the future long-baseline experiment ESSnuSB \cite{Baussan:2013zcy,Wildner:2015yaa}. The primary aim of the ESSnuSB experiment is to measure the Dirac CP-violating phase $\delta_{\rm CP}$ with high precision at the second oscillation maximum \cite{Ghosh:2019sfi}. Currently, there are two possible baseline options for ESSnuSB under consideration, which are 540~km and 360~km. In the present work, we consider both baseline options to estimate the sensitivity to invisible neutrino decay at ESSnuSB. The topics, which we address in this work, are the following: (i)~the capability of ESSnuSB to put bound on the decay parameter assuming there is no decay in Nature, (ii)~the capability of ESSnuSB to discover neutrino decay assuming that neutrinos decay in Nature, (iii)~how precisely ESSnuSB can measure the decay parameter if there exists neutrino decay in Nature, and (iv)~the effect of neutrino decay on the measurement of $\delta_{\rm CP}$. Since the sensitivity of ESSnuSB to measure the neutrino mass ordering, the octant of $\theta_{23}$, and the precision of $\theta_{23}$ is weak \cite{Agarwalla:2014tpa,Chakraborty:2017ccm,Chakraborty:2019jlv,Blennow:2019bvl}, we will not address such potential measurements in this work.

This paper is organized as follows. In Section~\ref{sec2}, we will discuss how the phenomenon of neutrino oscillations in the standard three-flavor neutrino oscillation scenario is altered in the presence of invisible neutrino decay. In Section~\ref{sec3}, we will present the experimental setup of ESSnuSB along with the simulation details. In Section~\ref{sec4}, we will present our results, and finally in Section~\ref{sec5}, we will summarize and conclude.

\section{Effects of invisible neutrino decay in neutrino oscillations}
\label{sec2}

In the standard three-flavor neutrino oscillation framework, the evolution equation can be written as
\begin{equation}
i \frac{d}{dt} \begin{pmatrix} \nu_e \\ \nu_\mu \\ \nu_\tau \end{pmatrix}
= \left\{ U \left[ \frac{1}{2E} {\rm diag}(0, \Delta m^2_{21}, \Delta m^2_{31})\right] U^\dagger + {\rm diag}(V, 0, 0)  \right\} \begin{pmatrix} \nu_e \\ \nu_\mu \\ \nu_\tau \end{pmatrix},
\label{eq:prop}
\end{equation}
where $U$ is the leptonic mixing matrix, $E$ is the neutrino energy, and $V = \sqrt{2}G_F n_e$ is the effective matter potential with $G_F$ being the Fermi coupling constant and $n_e$ the electron density of matter (along the neutrino trajectory). The sign of $V$ is positive for neutrinos and negative for antineutrinos. Note that the Hamiltonian in Eq.~(\ref{eq:prop}) is Hermitian and can be diagonalized by a unitary transformation. Assuming that the neutrino state $\nu_3$ decays into a sterile state, which effectively means that $\Delta m_{31}^2 \to \Delta m_{31}^2 - i \alpha_3$ \cite{Lindner:2001fx}, the evolution equation is modified to
\begin{equation}
i \frac{d}{dt} \begin{pmatrix} \nu_e \\ \nu_\mu \\ \nu_\tau \end{pmatrix}
= \left\{ U \left[ \frac{1}{2E} {\rm diag}\left(0, \Delta m^2_{21}, \Delta m^2_{31} - i \alpha_3 \right)\right] U^\dagger + {\rm diag}(V, 0, 0) \right\} \begin{pmatrix} \nu_e \\ \nu_\mu \\ \nu_\tau \end{pmatrix},
\end{equation}
where $\alpha_3 \equiv m_3/\tau_3$ with $\tau_3$ being the rest-frame lifetime of the neutrino state $\nu_3$ having mass $m_3$. We assume that the invisible neutrino state is lighter than the lightest active neutrino state, and therefore, decay is possible for both NO and IO. It is interesting to note that the Hamiltonian with decay is no longer Hermitian and therefore cannot be diagonalized by a unitary transformation. In this case, we follow the prescription given in Ref.~\cite{Hahn:2006hr} to numerically diagonalize this Hamiltonian to calculate the neutrino oscillation probabilities. For ESSnuSB, the impact of decay comes in both the appearance channel ($\nu_\mu \rightarrow \nu_e$) and the disappearance channel ($\nu_\mu \rightarrow \nu_\mu$). In vacuum, using the approximations $\Delta m_{21}^2 \sim 0$, $m_1 L/(\tau_1 E) \sim m_2 L/(\tau_2 E) \ll 1$, and $m_3 L/(\tau_3 E) \sim 1$, where $L$ is the baseline length, the neutrino oscillation probabilities relevant for ESSnuSB in presence of decay can be expressed as \cite{Lindner:2001fx,Giunti:2007ry,Akhmedov:2004ny}
\begin{align}
P_{\mu e} &\simeq s_{13}^2 c_{13}^2 s_{23}^2 \left[4 \sin^2 \frac{\Delta_{\rm atm}}{2} - \left(1 - e^{-\Gamma_3} \right) + 2 \cos\Delta_{\rm atm}\left(1 - e^{-\frac{\Gamma_3}{2}} \right)\right], \label{eq:Pme} \\
P_{\mu \mu} &\simeq 1 - c_{13}^2 s_{23}^2 \bigg[ 4 (1 - c_{13}^2 s_{23}^2) \sin^2 \frac{\Delta_{\rm atm}}{2} \nonumber\\
&+ c_{13}^2 s_{23}^2 \left(1 - e^{-\Gamma_3} \right) + 2 (1 - c_{13}^2 s_{23}^2) \cos\Delta_{\rm atm}\left(1 - e^{-\frac{\Gamma_3}{2}} \right) \bigg], \label{eq:Pmm}
\end{align}
where $s_{ij} \equiv \sin\theta_{ij}$, $c_{ij} \equiv \cos\theta_{ij}$, $\Delta_{\rm atm} \equiv \Delta m^2_{31} L/(2E)$, and $\Gamma_3 \equiv \alpha_3 L/E = m_3 L/(\tau_3 E)$. Note that as Eq.~(\ref{eq:Pme}) is derived using the approximation $\Delta m_{21}^2 \sim 0$, there is no $\delta_{\rm CP}$ term in this equation. This is because the parameter $\delta_{\rm CP}$ appears in the appearance channel probability from the interference term between $\Delta m_{21}^2$ and $\Delta m_{31}^2$ and we set $\Delta m_{21}^2=0$ in the above for simplicity.  From Eqs.~({\ref{eq:Pme}) and (\ref{eq:Pmm}), we understand that invisible neutrino decay leads to a depletion in the number of events in both appearance and disappearance channels. We also note that the decay parameter appears together with the quantity $L/E$. Since the neutrino energy $E$ is same for both baseline options for ESSnuSB, which are $L = 540$~km and $L = 360$~km,  we expect to observe different effects of decay for different baseline lengths.

We must point out that while the approximate neutrino oscillation probabilities given in Eqs.~({\ref{eq:Pme}) and (\ref{eq:Pmm}) have been derived to illustrate the impact of decay on neutrino oscillations, all results presented in the following sections have been obtained from a full numerical simulation of the neutrino propagation equations in matter taking into account full three-flavor effects. We have explicitly checked that the approximate expression of the probability for the disappearance channel given in Eq.~(\ref{eq:Pmm}) matches with the exact numerical results to a very good accuracy. On the other hand, for the appearance channel, the approximate expression given in Eq.~(\ref{eq:Pme}) has a mismatch with the exact probability, since as was pointed out before, we have neglected $\Delta m_{21}^2$ in this approximation. We will observe in Section~\ref{sec4} that the sensitivity of ESSnuSB to invisible neutrino decay comes essentially from the disappearance channel, while the significance of the appearance channel is marginal. Therefore, in the rest of this section, we will only look into the disappearance channel, using the approximate probability given in Eq.~(\ref{eq:Pmm}) in order to understand the impact of decay.

\begin{figure}
\begin{center}
\includegraphics[width=0.49\textwidth]{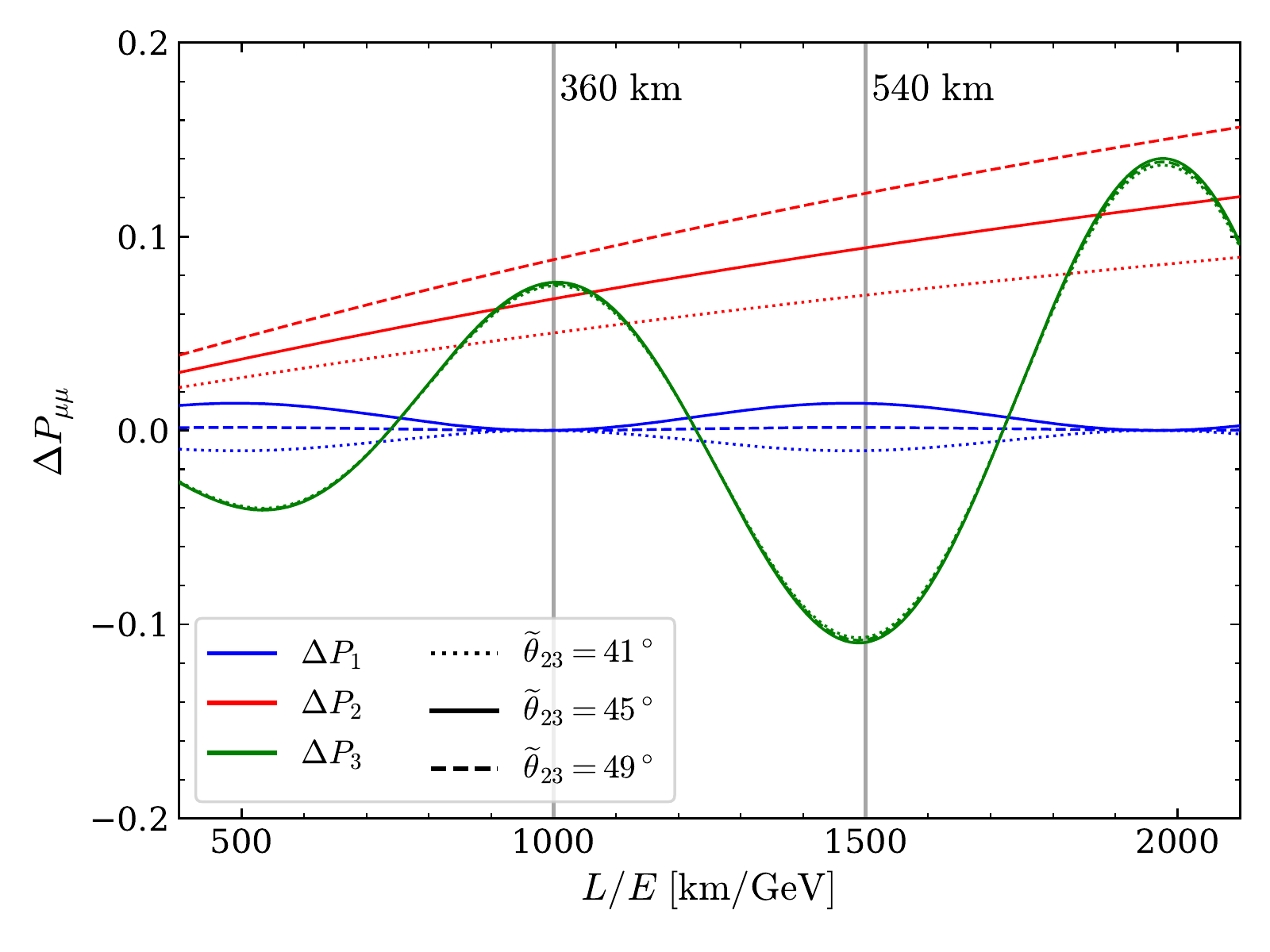} 
\end{center}
\vspace{-5mm}
\caption{Difference in neutrino oscillation probability $\Delta P_{\mu \mu} = P_{\mu \mu}^{\mathrm{std}}(\theta_{23}=49.2^\circ) - P_{\mu \mu}^{\mathrm{decay}}(\widetilde{\theta}_{23})$ between the standard oscillation probability, $P_{\mu \mu}^{\mathrm{std}}$, and the probability with invisible decay, $P_{\mu \mu}^{\mathrm{decay}}$, using the decay parameter $\tau_3/m_3 = 1.0 \times 10^{-11}$~s/eV. The probabilities are calculated using Eq.~(\ref{eq:Pmm}). In the probability including decay, the value of the mixing angle $\widetilde{\theta}_{23}$ is plotted for three different values: $41^\circ$, $45^\circ$, and $49^\circ$ with dotted, solid, and dashed curves, respectively. The other neutrino oscillation parameters used are listed in Table~\ref{Tab:Param}. The three terms $\Delta P_1$, $\Delta P_2$, and $\Delta P_3$ are displayed in blue, red, and green, respectively. The grey vertical lines mark the flux peak at $0.36$~GeV for the two baseline options of 360~km and 540~km.}
\label{fig_terms}
\end{figure}

Since the main aim of the current work is to determine how well ESSnuSB can distinguish the standard case from the decay case, we present in Fig.~\ref{fig_terms} the probability difference
\begin{equation}
\Delta P_{\mu \mu} \equiv P_{\mu \mu}^{\mathrm{std}}(\theta_{23}=49.2^\circ) - P_{\mu \mu}^{\mathrm{decay}}(\widetilde{\theta}_{23}),
\label{eq:DeltaPmm}
\end{equation}
where $P_{\mu \mu}^{\mathrm{std}}$ is the neutrino oscillation probability in the standard case without decay using the mixing angle $\theta_{23} = 49.2^\circ$ and $P_{\mu \mu}^{\mathrm{decay}}$ is the probability as a function of the effective mixing angle $\widetilde{\theta}_{23}$ with invisible decay using the decay parameter $\tau_3/m_3 = 1.0 \times 10^{-11}$~s/eV. We divide the probability difference $\Delta P_{\mu \mu}$ in Eq.~(\ref{eq:DeltaPmm}) as follows
\begin{equation}
\Delta P_{\mu \mu} = \Delta P_1  +  \Delta P_2   +  \Delta P_3,
\end{equation}
where we define the three terms
\begin{align}
\Delta P_1 &\equiv 4 c_{13}^2 \left[ \widetilde{s}_{23}^2 (1 - c_{13}^2 \widetilde{s}_{23}^2) - s_{23}^2 (1 - c_{13}^2 s_{23}^2) \right] \sin^2 \frac{\Delta_{\rm atm}}{2}, \\
\Delta P_2 &\equiv c_{13}^4 \widetilde{s}_{23}^4 \left(1 - e^{-\Gamma_3} \right), \\
\Delta P_3 &\equiv 2 c_{13}^2 \widetilde{s}_{23}^2 (1 - c_{13}^2 \widetilde{s}_{23}^2) \cos\Delta_{\rm atm}\left(1 - e^{-\frac{\Gamma_3}{2}} \right).
\end{align}
In Fig.~\ref{fig_terms}, each of these terms are shown for three different values of $\widetilde\theta_{23}$, i.e., $41^\circ$ (dotted curves), $45^\circ$ (solid curves), and $49^\circ$ (dashed curves). The purpose of this figure is mainly to show the effect of $\theta_{23}$ on the difference between the oscillation probability for no decay and decay. Note that $\Delta P_2$ and $\Delta P_3$ are zero in the standard case with no decay, whereas $\Delta P_1$ does not depend on the decay parameter. Therefore, since $\Delta P_2$ and $\Delta P_3$ are equal to zero if $\Gamma_3 = 0$ (i.e., there is no decay), the expressions for $\Delta P_2$ and $\Delta P_3$ are independent of $\theta_{23}$, but not of $\widetilde\theta_{23}$. We will use this figure in the later sections to explain our main results on the sensitivity of ESSnuSB to the decay parameter. We can already note a few points from the figure. The term $\Delta P_1$ depends on $\sin^2 (\Delta_{\rm atm}/2)$ and is hence observed to be oscillating with a small amplitude as a function of $L/E$, since it is proportional to the difference $4 c_{13}^2 \left[ \widetilde{s}_{23}^2 (1 - c_{13}^2 \widetilde{s}_{23}^2) - s_{23}^2 (1 - c_{13}^2 s_{23}^2) \right]$, where $s_{23}$ and $\widetilde{s}_{23}$ are taken to be very close to each other. The reason is that it does not depend on the decay parameter and the small non-zero oscillatory value it obtains (roughly 3~\%) comes from the fact that $\theta_{23}=49.2^\circ$, while $\widetilde{\theta}_{23}$ is assumed to be $41^\circ$, $45^\circ$, and $49^\circ$, for the three cases, respectively. On the other hand, we see that the terms $\Delta P_2$ and $\Delta P_3$ have rather non-trivial behaviors and values. The term $\Delta P_2$ is independent of $\Delta_{\rm atm}$, and hence is non-oscillatory. It increases almost linearly with $L/E$ as $\Delta P_2 \simeq c_{13}^4 \widetilde{s}_{23}^4 \alpha_3 L/E$ for $\Gamma_3$ small. The spread in $\Delta P_2$ with $\widetilde{s}_{23}$ can be observed in the figure. In contrast, the term $\Delta P_3$ is observed to be oscillating [with a relatively large amplitude proportional to $2 c_{13}^2 \widetilde{s}_{23}^2 (1 - c_{13}^2 \widetilde{s}_{23}^2$)] as a function of $L/E$, since it depends on $\cos \Delta_{\rm atm}$. We show by the two grey vertical lines the $L/E$ corresponding to the 360~km and 540~km baseline options, respectively,} where we take $E=0.36$~GeV, which is the energy where the ESSnuSB flux peaks. Note from the figure that for the 360~km baseline option, $\Delta P_3$ has a peak, whereas for the 540~km baseline option, it has a trough. Contrast this with the behavior of $\Delta P_2$, which is increasing almost linearly and is positive for both the 360~km and 540~km baseline options. This means that while $\Delta P_2$ and $\Delta P_3$ add up for the 360~km baseline option, they cancel each other for the 540 km one.  We will see that this has far reaching consequences when it comes to marginalization over $\theta_{23}$.

\section{Experimental setup and simulation details of ESSnuSB}
\label{sec3}

We use the software GLoBES \cite{Huber:2004ka,Huber:2007ji} to simulate the sensitivity of the ESSnuSB experiment. We consider a water-\cnv\ far detector \cite{Agostino:2012fd} of fiducial volume 507~kt located at distance of either 540~km or 360~km from the neutrino source. We also consider an identical near detector located at a distance of 0.5~km having volume of 0.1~kt. For the neutrino source, we consider protons of 2.5~GeV originating from a beam of 5~MW capable of delivering $2.7\times 10^{23}$ protons on target  per year. We assume a total run-time of 10~years divided into 5~years in neutrino mode and 5~years in antineutrino mode. We consider correlated systematics between far and near detectors with the errors as given in Ref.~\cite{Coloma:2012ji} and we list them in Table~\ref{Tab:Systematics} for convenience.
\begin{table}
\centering
\begin{tabular} {| c | c |}
\hline
Systematics &  Default \\
\hline
Fiducial volume ND & 0.5~\% \\
Fiducial volume FD & 2.5~\% \\
Flux error $\nu$ & 7.5~\% \\
Flux error $\bar{\nu}$ & 15~\% \\
Neutral current background & 7.5~\% \\
Cross section $\times$ eff. QE & 15~\% \\
Ratio $\nu_e/\nu_{\mu}$ QE & 11~\% \\
\hline
\end{tabular}
\caption{Systematic uncertainties for a super beam as described in Ref.~\cite{Coloma:2012ji} for the ``Default'' scenario.}
\label{Tab:Systematics}
\end{table}

We estimate the statistical $\chi^2$ function using
\begin{equation}
\chi^2_{{\rm stat}} = 2 \sum_{i=1}^n \bigg[ N^{{\rm test}}_i - N^{{\rm true}}_i - N^{{\rm true}}_i \log\bigg(\frac{N^{{\rm test}}_i}{N^{{\rm true}}_i}\bigg) \bigg]\,,
\end{equation}
where $n$ is the number of energy bins, $N^{{\rm true}}$ is the number of true events, and $N^{{\rm test}}$ is the number of test events and incorporate the systematics by the method of pulls. Unless otherwise mentioned, in our simulation, we generate the data with the best-fit values from the global analysis of the world neutrino data as obtained by NuFIT~v5.0 \cite{Esteban:2020cvm} and we present them in Table~\ref{Tab:Param}.
\begin{table}
	\centering
	\setlength{\extrarowheight}{0.1cm}
	\begin{tabular}{|c|c|c|}
		\hline
		 Parameter & Best-fit value & $3 \sigma$ allowed values\\
		\hline
		$\theta_{12}$ & $33.44^\circ$ & $(31.27^\circ, 35.86^\circ)$\\
		$\theta_{13}$ & $8.57^\circ$ & $(8.20^\circ, 8.93^\circ)$\\
		$\theta_{23}$ & $49.2^\circ$ & $(40.1^\circ, 51.7^\circ)$\\
		$\delta_{\rm CP}$ & $-163^\circ$ & $(-180^\circ, 9^\circ)$ and $(120^\circ, 180^\circ)$\\
		$\Delta m^2_{21}$ & $7.42\times10^{-5}~\mathrm{eV}^2$ & $(6.82, 8.04)\times10^{-5}~\mathrm{eV}^2$ \\
		$\Delta m^2_{31}$ & $2.517\times10^{-3}~\mathrm{eV}^2$ & $(2.435, 2.598)\times10^{-3}~\mathrm{eV}^2$\\
		\hline
	\end{tabular}
	\caption{Best-fit values of the neutrino oscillation parameters and the corresponding $3 \sigma$ allowed values for the standard three-flavor scenario. The values have been adopted from Ref.~\cite{Esteban:2020cvm}.}
	\label{Tab:Param}
\end{table}
In the fit, we minimize over the parameters $\theta_{13}$, $\theta_{23}$, and $\delta_{\rm CP}$ in their current $3 \sigma$ ranges, as given in Table~\ref{Tab:Param}. We keep the parameters $\theta_{12}$, $\Delta m^2_{21}$, and $\Delta m^2_{31}$ fixed to their best-fit values in the fit. Throughout our analysis, we assume that the mass ordering of the neutrinos is known and is NO.

\section{Simulation results}
\label{sec4}

We discuss the sensitivity of ESSnuSB in presence of invisible neutrino decay. Our strategy is as follows. First, we show how the neutrino oscillation probabilities in the $\nu_\mu \rightarrow \nu_e$ and $\nu_\mu \rightarrow \nu_\mu$ channels are modified due to the presence of decay for both baseline options of ESSnuSB. We also show the event rates without decay, to demonstrate the relevant energy values from where the sensitivity stems. Then, we study the capability of ESSnuSB to put bound on the decay parameter and compare our results with other experimental setups. We also present the discovery potential of ESSnuSB to observe neutrino decay. Next, we study how precisely ESSnuSB can measure the decay parameter if decay exists in Nature. Finally, we discuss the effects of decay in CP violation discovery and CP precision measurements at ESSnuSB.

\subsection{Discussion at the probability and event level}
\label{Sec:ProbLevel}

\begin{figure}
\begin{center}
\includegraphics[width=0.49\textwidth]{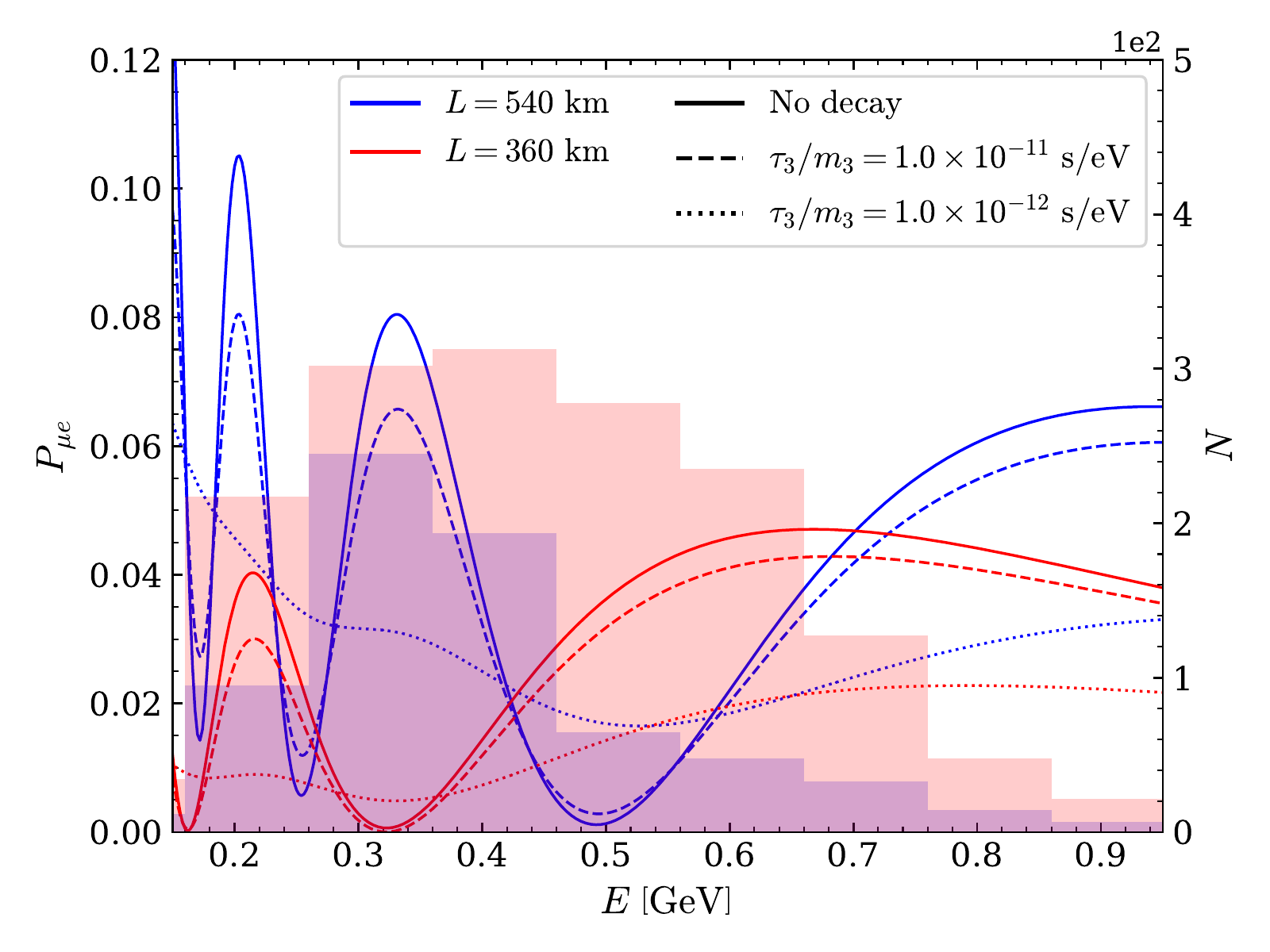} 
\includegraphics[width=0.49\textwidth]{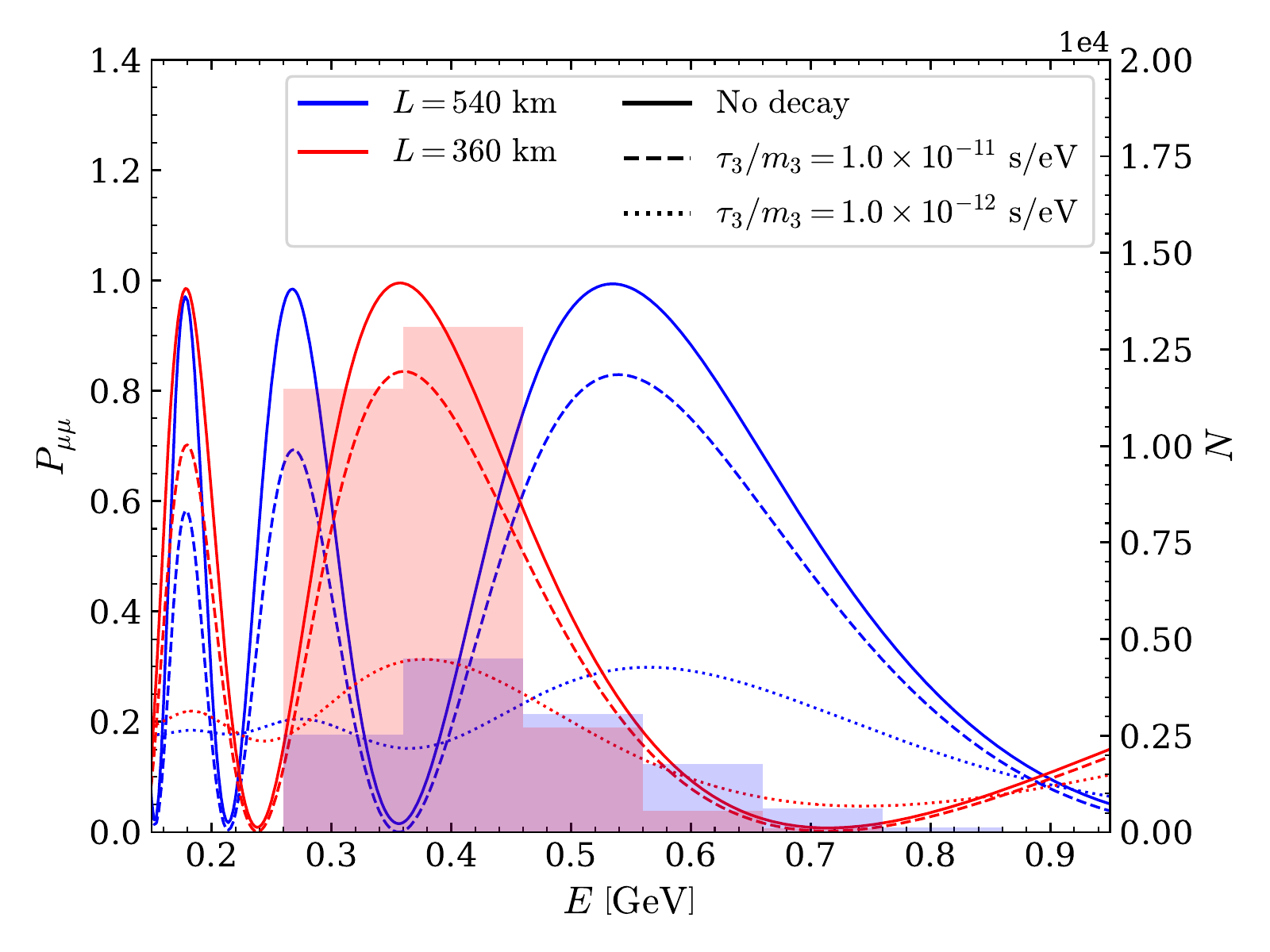} 
\end{center}
\vspace{-5mm}
\caption{Neutrino oscillation probabilities and neutrino event rates as functions of neutrino energy $E$ for both baseline options of ESSnuSB. The event rates represented by the histograms are without decay.}
\label{fig_prob}
\end{figure}

In Fig.~\ref{fig_prob}, we present the neutrino oscillation probabilities and neutrino event rates of ESSnuSB as functions of neutrino energy $E$.
The left panel is for the appearance channel, whereas the right panel is for the disappearance channel. In each panel, the blue curve (histogram) corresponds to the probability (event rate) for $L = 540$~km and the red curve (histogram) corresponds to the probability (event rate) for $L = 360$~km. The solid curves correspond to the case when there is no decay. The histograms are plotted for the no decay scenario. To generate these curves, we use the best-fit values of the standard neutrino oscillation parameters given in Table~\ref{Tab:Param}. To understand the effect of decay in the neutrino oscillation probabilities, we consider two values of the decay parameter, i.e., $\tau_3/m_3 = 10^{-11}$~s/eV and $\tau_3/m_3 = 10^{-12}$~s/eV, and they are shown by dashed and dotted curves, respectively. In the appearance channel (the left panel of Fig.~\ref{fig_prob}), we realize that for $L = 540$~km, most of the sensitivity is due to the second oscillation maximum. In fact, for $L = 360$~km, the sensitivity comes from both the first and second oscillation maxima. In the disappearance channel (the right panel of Fig.~\ref{fig_prob}), we note that for $L = 540$~km, the sensitivity actually comes from the second oscillation minimum, whereas for $L = 360$~km, it stems from the second oscillation maximum. From the figure, we also observe that the effect of decay is non-negligible in both the appearance and disappearance channels. Between $\tau_3/m_3 = 10^{-11}$~s/eV and $\tau_3/m_3 = 10^{-12}$~s/eV, the separation of the no decay curve and the decay curve is larger for $\tau_3/m_3 = 10^{-12}$~s/eV, reflecting the fact that the change in the probability is much larger for a shorter lifetime of the neutrino state. Note that the number of events are higher for $L = 360$~km as compared to $L = 540$~km, since the number events are proportional to  $1/L^2$.

\subsection{Sensitivity and discovery potential in presence of invisible neutrino decay}
\label{Bounds}

In Fig.~\ref{fig_sens}, we plot the capability of ESSnuSB to put bound on the decay parameter (or the sensitivity $\chi^2$) in the left panel and its potential to discover decay (or the discovery $\chi^2$) in the right panel.
\begin{figure}
\begin{center}
\includegraphics[width=0.49\textwidth]{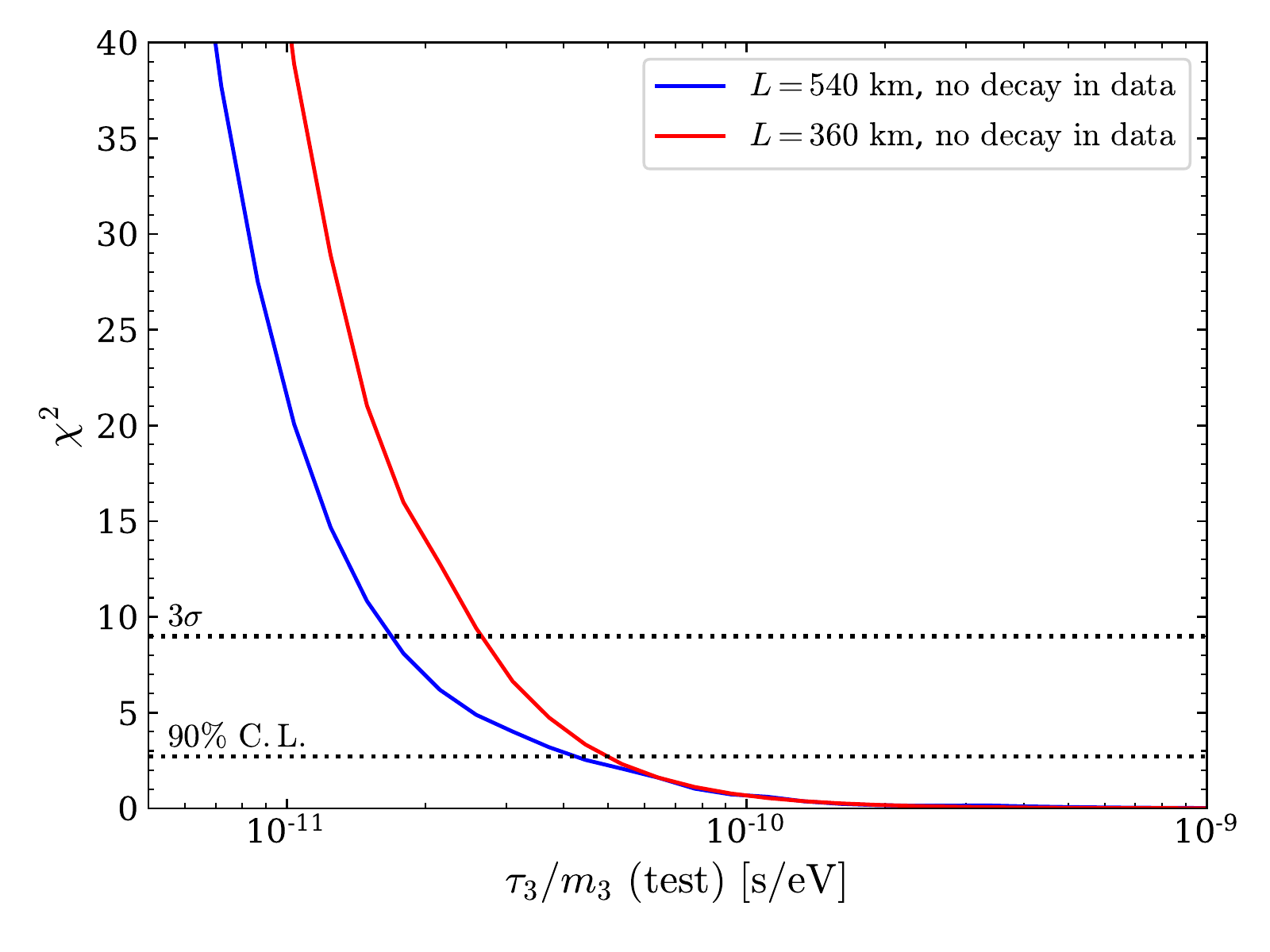} 
\includegraphics[width=0.49\textwidth]{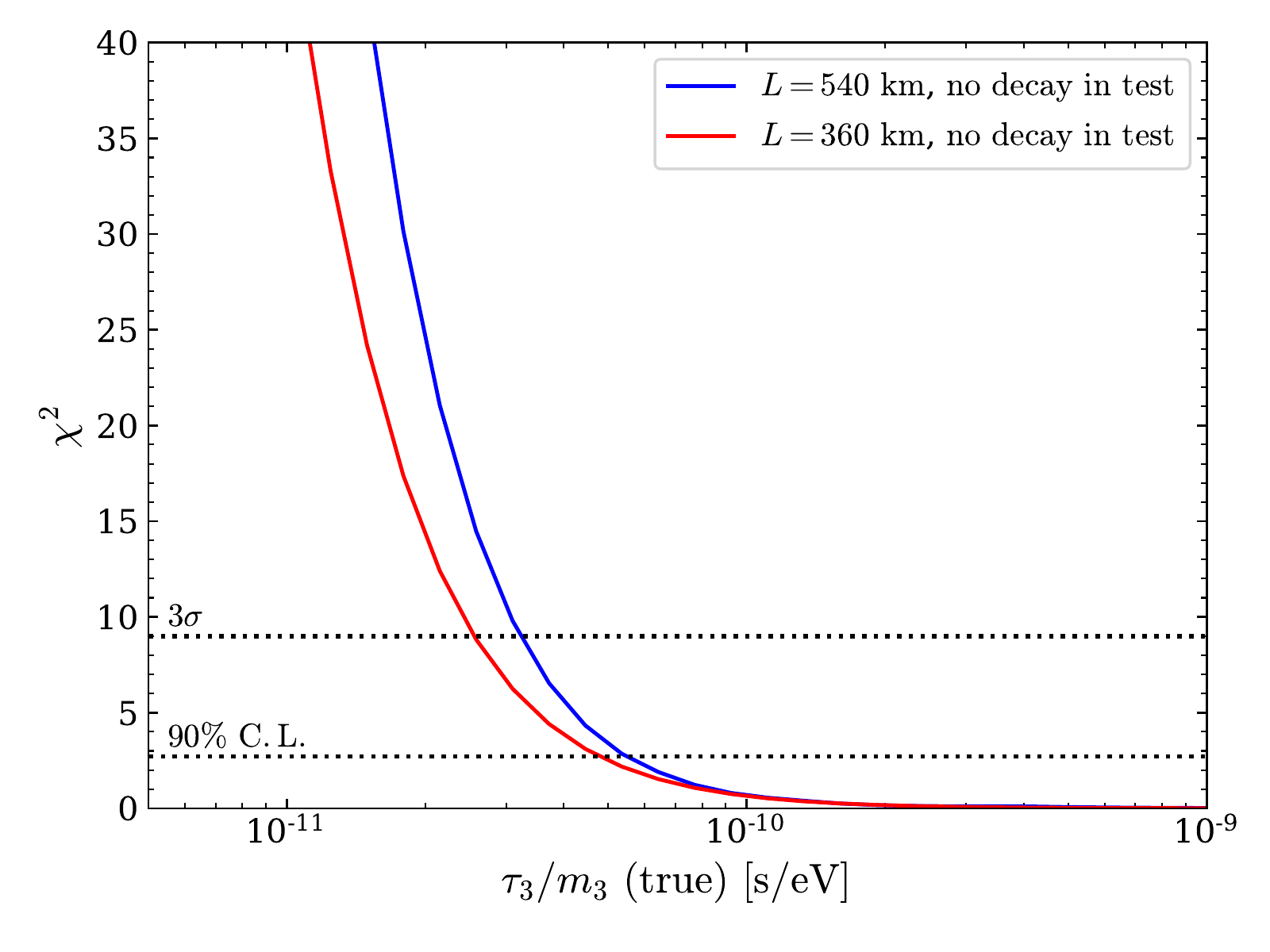} 
\end{center}
\vspace{-5mm}
\caption{Sensitivity $\chi^2$ (left panel) and discovery $\chi^2$ (right panel) as functions of $E$ for both baseline options of ESSnuSB.}
\label{fig_sens}
\end{figure}

In the left panel, we have not considered decay in data, whereas in the right panel, we have not considered decay in theory. The black dotted horizontal lines represent the value of the $\chi^2$ corresponding to $90~\%$~C.L and $3 \sigma$. In each panel, the blue curve represents the sensitivity for $L = 540$~km and the red curve represents the sensitivity for $L = 360$~km. From the left panel, we observe that the bound on the decay parameter is $1.68 \times 10^{-11}$~s/eV for $L = 540$~km and $2.64 \times 10^{-11}$~s/eV for $L = 360$~km at $3 \sigma$. On the other hand, ESSnuSB can discover decay at $3 \sigma$ if  the true value of the decay parameter is $2.56 \times 10^{-11}$ s/eV for $L = 360$~km and $3.22 \times 10^{-11}$ s/eV for $L = 540$~km. We note that the sensitivity $\chi^2$ is better for the 360~km baseline option of ESSnuSB, whereas the discovery $\chi^2$ is better for the 540~km baseline option of ESSnuSB. To understand this result, we calculate the contribution to the $\chi^2$ function from the individual appearance and disappearance channels for both 540~km and 360~km and for both sensitivity and discovery. We find that the most of the contribution comes from the disappearance channel, which is plotted in Fig.~\ref{fig_disap}. 

\begin{figure}
\begin{center}
\includegraphics[width=0.49\textwidth]{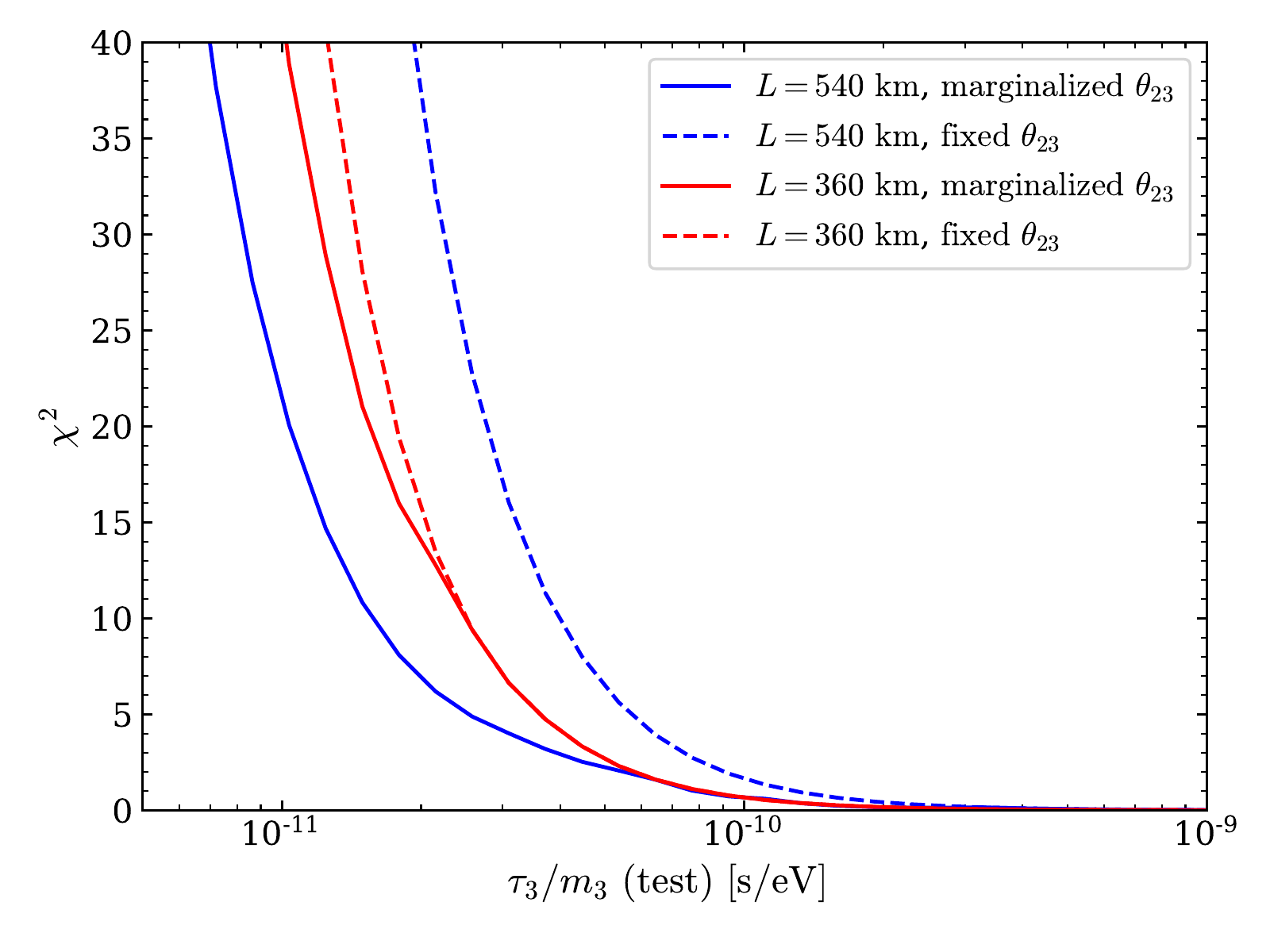} 
\includegraphics[width=0.49\textwidth]{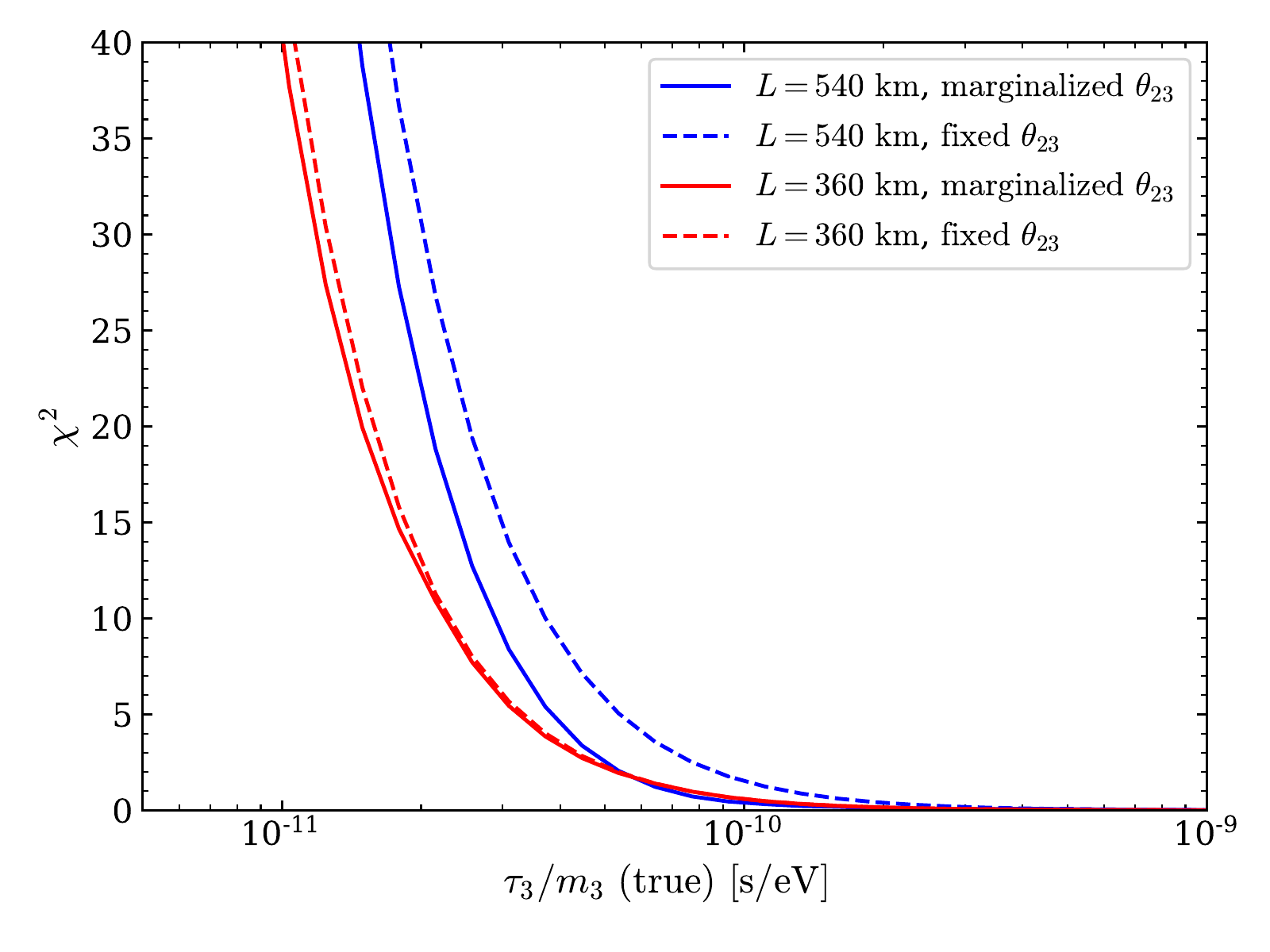} 
\end{center}
\vspace{-5mm}
\caption{Contribution from the disappearance channel to the sensitivity $\chi^2$ (left panel) and contribution from the disappearance channel to the discovery $\chi^2$ (right panel) as functions of $E$ for both baseline options of ESSnuSB.}
\label{fig_disap}
\end{figure}

\begin{figure}
\begin{center}
\includegraphics[width=0.49\textwidth]{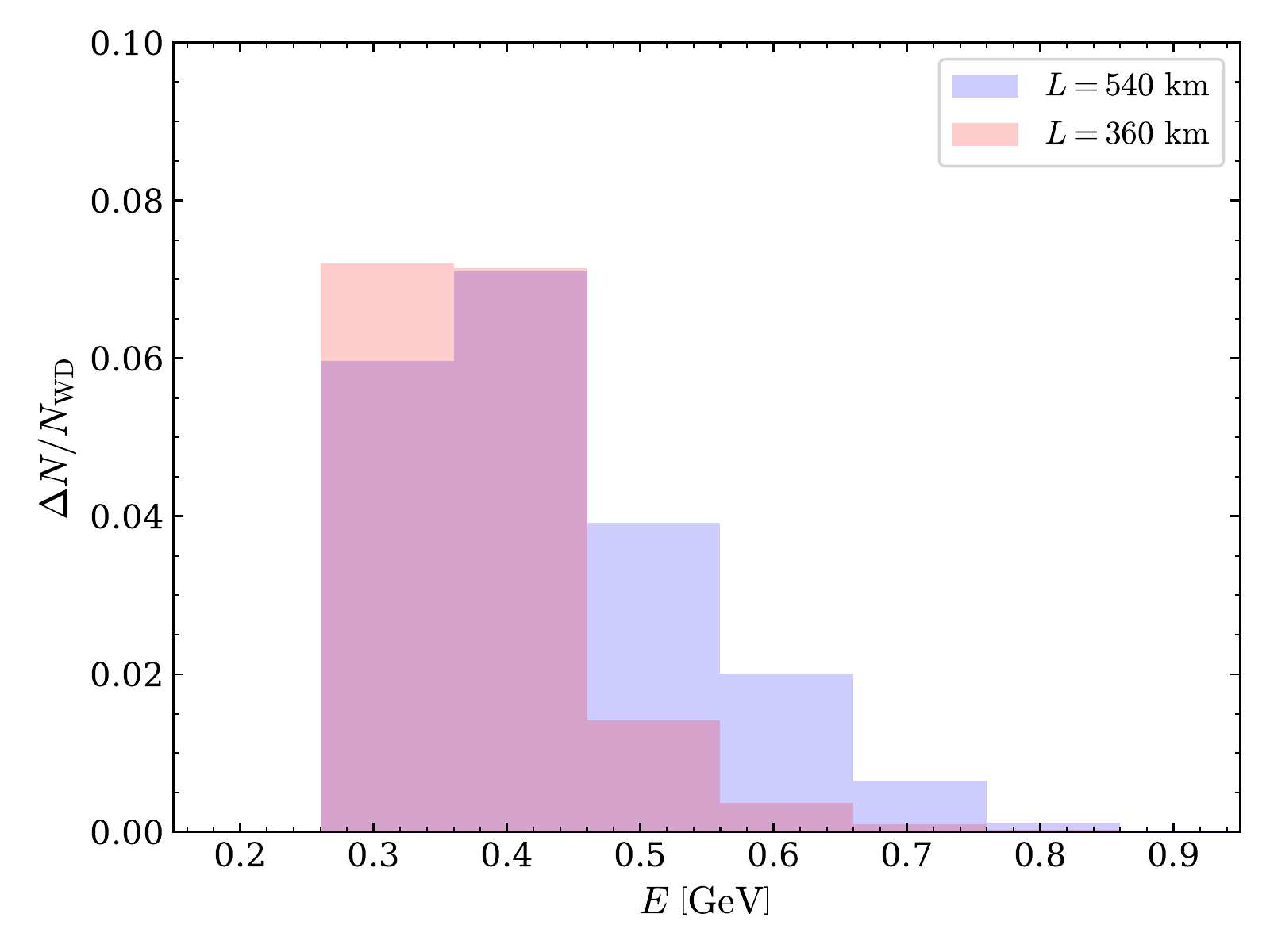} 
\includegraphics[width=0.49\textwidth]{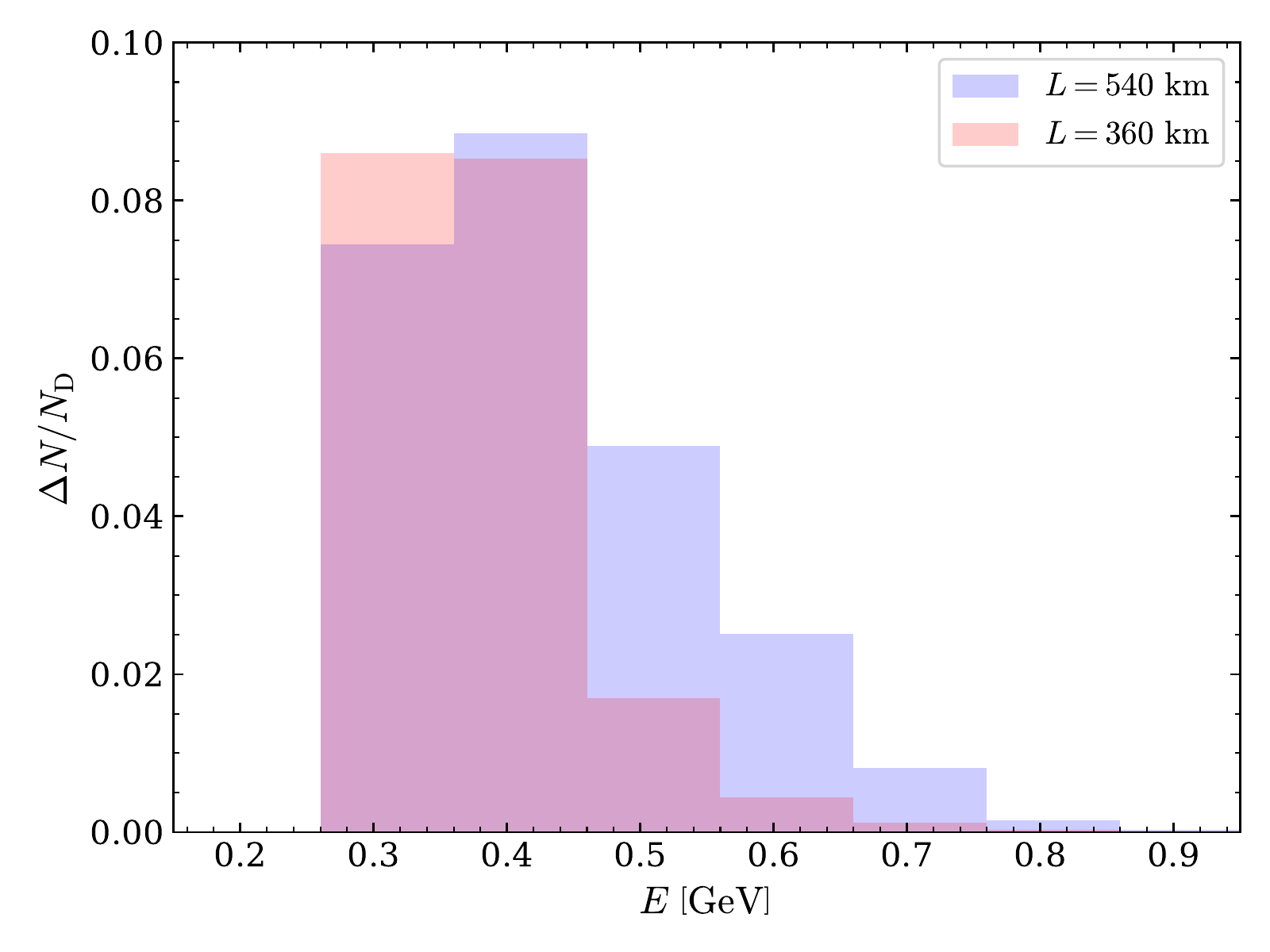} 
\end{center}
\vspace{-5mm}
\caption{Difference between the number of events without decay and the number of events with decay divided by the number of events with decay (left panel) and difference between the number of events without decay and the number of events with decay divided by the number of events without decay (right panel) for the disappearance channel as functions of $E$. The value of $\tau_3/m_3 = 10^{-11}$~s/eV.}
\label{fig_diff}
\end{figure}

In Fig.~\ref{fig_disap}, as was the case in Fig.~\ref{fig_sens}, the left panel is for the sensitivity $\chi^2$ and the right panel is for the discovery $\chi^2$.
The blue curves correspond to $L = 540$~km and red curves correspond to $L = 360$~km. In each panel, the solid curves correspond to the case when the $\chi^2$ is minimized over the parameter $\theta_{23}$ in theory and the dashed curves correspond to the case when the parameter $\theta_{23}$ is kept fixed to its best-fit value in theory. From the panels, we note that when $\theta_{23}$ is fixed, the 540~km baseline option of ESSnuSB is better than the 360~km baseline option, for both sensitivity and discovery. However, when the $\chi^2$ is minimized over the parameter $\theta_{23}$, we find that the $L = 360$~km option is better for sensitivity, whereas $L = 540$~km is better for discovery. The reason is that the effect of minimizing over $\theta_{23}$ is significant for the sensitivity $\chi^2$ for the $L = 540$~km option. In order to understand this, let us first see why the $L = 540$~km option is better for both sensitivity and discovery when $\theta_{23}$ is kept fixed.

In Fig.~\ref{fig_diff}, we display as a function of $E$ the difference between the number of events without decay and the number of events with decay divided by the number of events with decay, i.e., $\Delta N/N_{\rm D}$ (left panel), and the difference between the number of events without decay and the number of events with decay divided by the number of events without decay, i.e., $\Delta N/N_{\rm WD}$ (right panel) for the disappearance channel.
We generate these panels for the best-fit values of the neutrino oscillation parameters given in Table~\ref{Tab:Param} and the value of the decay parameter as $\tau_3/m_3 = 10^{-11}$~s/eV. The blue histograms are for $L = 540$~km and red histograms are for $L = 360$~km. Therefore, this figure reflects the contribution of each energy bin to the sensitivity (left panel) and discovery (right panel) $\chi^2$. From the figure, we observe that in the energy bin of 0.3~GeV and 0.4~GeV, $\Delta N/N_{\rm D}$ and $\Delta N/N_{\rm WD}$ are similar for both $L = 540$~km and $L = 360$~km. However, in the other energy bins, $\Delta N/N_{\rm D}$ and $\Delta N/N_{\rm WD}$ are higher for $L = 540$~km as compared to $L = 360$~km. This explains why for a fixed value of $\theta_{23}$, both the sensitivity and discovery $\chi^2$ are better for $L = 540$~km. Next, let us understand what is the effect of minimization of  the sensitivity $\chi^2$ over $\theta_{23}$ for the $L = 540$~km option.

This can be understood from Fig.~\ref{fig_th23}, where we plot the disappearance channel probability for $L = 540$~km (left panel) and $L = 360$~km (right panel) as functions of $E$.
\begin{figure}
\begin{center}
\includegraphics[width=0.49\textwidth]{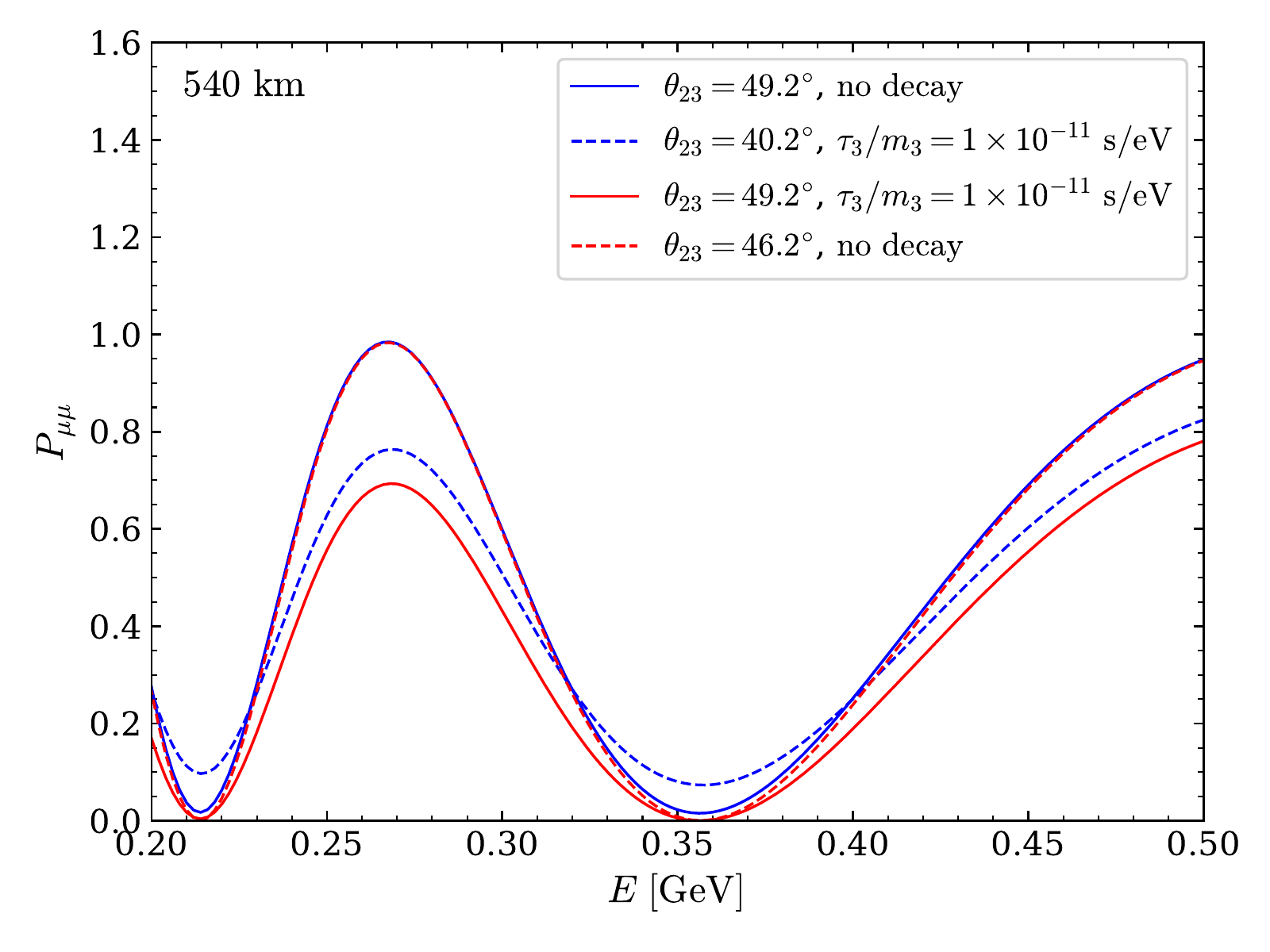} 
\includegraphics[width=0.49\textwidth]{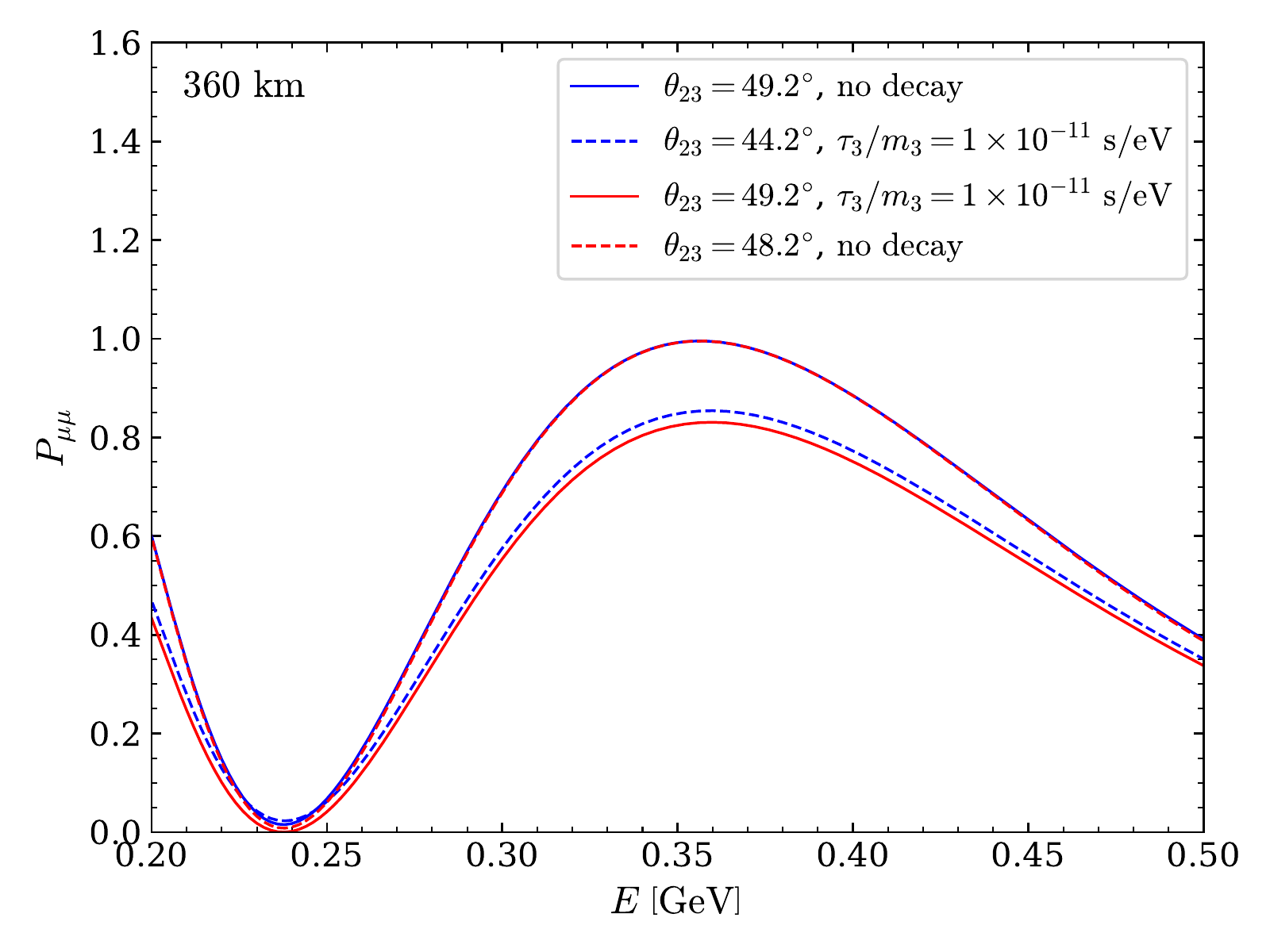} 
\end{center}
\vspace{-5mm}
\caption{Disappearance channel probability for $L = 540$~km (left panel) and $L = 360$~km (right panel) as functions of $E$, considering decay and no decay for different values of $\theta_{23}$.}
\label{fig_th23}
\end{figure}
The presented curves are for different values of $\theta_{23}$, corresponding to either the true value of $\theta_{23}$ or the value of $\theta_{23}$ where the $\chi^2$ minimum is occurring, considering both decay and no decay. The value of the decay parameter is $\tau_3/m_3 = 10^{-11}$~s/eV. From Fig.~\ref{fig_prob}, we understand that for the disappearance channel the sensitivity comes from $E \sim 0.4$ GeV. Therefore, in these panels, we focus on the values of the probability around $E \sim 0.4$~GeV. For $L = 540$~km, we note that the separation between the blue solid curve (which is the true point for the sensitivity $\chi^2$) and the red solid curve (which is the test point for the sensitivity $\chi^2$ for fixed $\theta_{23}$) is larger than the separation between the blue solid curve and the blue dashed curve (which is the test point for the sensitivity $\chi^2$ when $\theta_{23}$ is minimized). On the other hand, for $L = 360$~km, the separation between the blue solid curve and the red solid curve is similar to the separation between the blue solid curve and the blue dashed curve. For this reason, the effect of $\theta_{23}$ marginalization is larger for the sensitivity $\chi^2$ for $L = 540$~km. Similarly, for the discovery $\chi^2$, one can see that the separation between the red solid curve (which is the true point for the discovery $\chi^2$) and the blue solid curve (which is the test point for the discovery $\chi^2$ for fixed $\theta_{23}$) and the separation between the red solid curve and the red dashed curve (which is the test point for the discovery $\chi^2$ when $\theta_{23}$ is minimized) is very similar for both $L = 540$~km and $L = 360$~km. This is why the discovery $\chi^2$ is not affected by the $\theta_{23}$ minimization. 

To see this even more clearly, let us go back to our approximate expression for the survival probability given in Eq.~(\ref{eq:Pmm}) and the related figure, i.e., Fig.~\ref{fig_terms}. For the sensitivity study, the data are generated for no decay and we assume $\theta_{23} =49.2^\circ$. This is then fitted with a theory with decay, while allowing $\theta_{23}$ to vary within its current $3\sigma$ bound. The fit, of course, tries to minimize the difference between the data and the theory -- this essentially means that $\Delta P_{\mu\mu}$ is reduced to the best possible case in the fit. If we now look at the three terms plotted in Fig.~\ref{fig_terms} for the three test values of $\widetilde\theta_{23}$, we note the following. The $\theta_{23}$ dependence is mostly coming from $\Delta P_2$ and for $\widetilde\theta_{23}=49^\circ$, the theory is worst as compared to the data. The difference between data and theory (i.e., $\Delta P_{\mu\mu}$) starts to reduce as we reduce $\theta_{23}$ in the fit. Note that the spread in $\Delta P_{\mu\mu}$ due to $\Delta P_2$ coming from $\theta_{23}$ is larger for the 540~km baseline option as compared to the 360~km one. Therefore, it is expected that the effect of $\theta_{23}$ marginalization will be larger for the 540~km baseline option. Indeed, this is what we have observed in Fig.~\ref{fig_terms}. In addition, there is yet another point to note that brings a difference between the 360~km and 540~km baseline options. If we look at $\Delta P_3$, we see that while this term itself is largely independent of $\theta_{23}$, it has a very interesting correlation with respect to $\Delta P_2$, and that makes it relevant in $\theta_{23}$ marginalization. The term $\Delta P_3$ is oscillatory and we can see from Fig.~\ref{fig_terms} that while it has a peak for the 360~km baseline option, it has a trough for the 540~km baseline option. This results in partial cancellation between $\Delta P_2$ and $\Delta P_3$ for the 540~km option. Since $\Delta P_2$ depends on $\theta_{23}$, the extent of this cancellation depends on the value of $\theta_{23}$ and as a result the $\chi^2$ for the 540~km option becomes lower as a result of marginalization over $\theta_{23}$. 

A similar argument can be used to understand how $\theta_{23}$ marginalization affects the $\chi^2$ for discovery and why for that case too the effect is larger for the 540~km baseline option. However, there is a major difference. The data are now generated for decay, while the fit is performed for standard oscillations. In Fig.~\ref{fig_terms}, this would mean that $\Delta P_{\mu\mu}$ would flip sign for all three terms, but this is not the main issue. The more important difference is that now we have standard oscillations in the fit. This means that $\Delta P_2$ and $\Delta P_3$ are absent in the fit and only $\Delta P_1$ can be manipulated to minimize the $\chi^2$. Since $\Delta P_1$ has only a small dependence on $\theta_{23}$, the effect of marginalization on the discovery $\chi^2$ is smaller. The difference in $\theta_{23}$ marginalization between the 360~km and 540~km baseline options can also be seen from the same figure. We observe that for the 360~km baseline option, $\Delta P_1$ is almost zero, and hence independent of $\theta_{23}$. Therefore, there is a small effect of marginalization on the discovery $\chi^2$ for this case. For the 540~km baseline option, the effect is larger, since for this case, there is a peak for $\Delta P_1$. However, even for this baseline option, the $\theta_{23}$ marginalization effect is significantly smaller as compared to the sensitivity case for the reason mentioned above. 

\begin{figure}
\begin{center}
\includegraphics[width=0.49\textwidth]{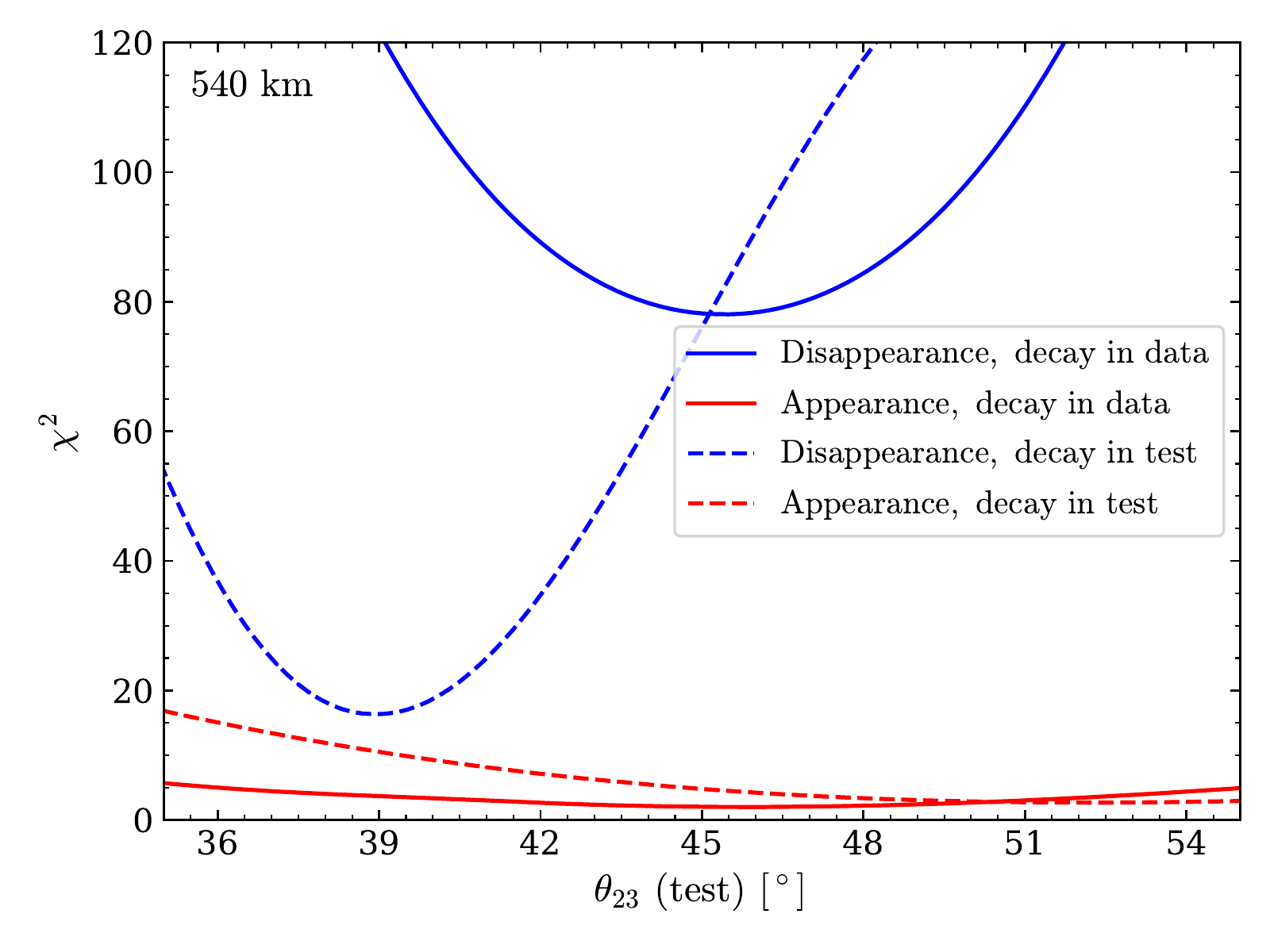} 
\includegraphics[width=0.49\textwidth]{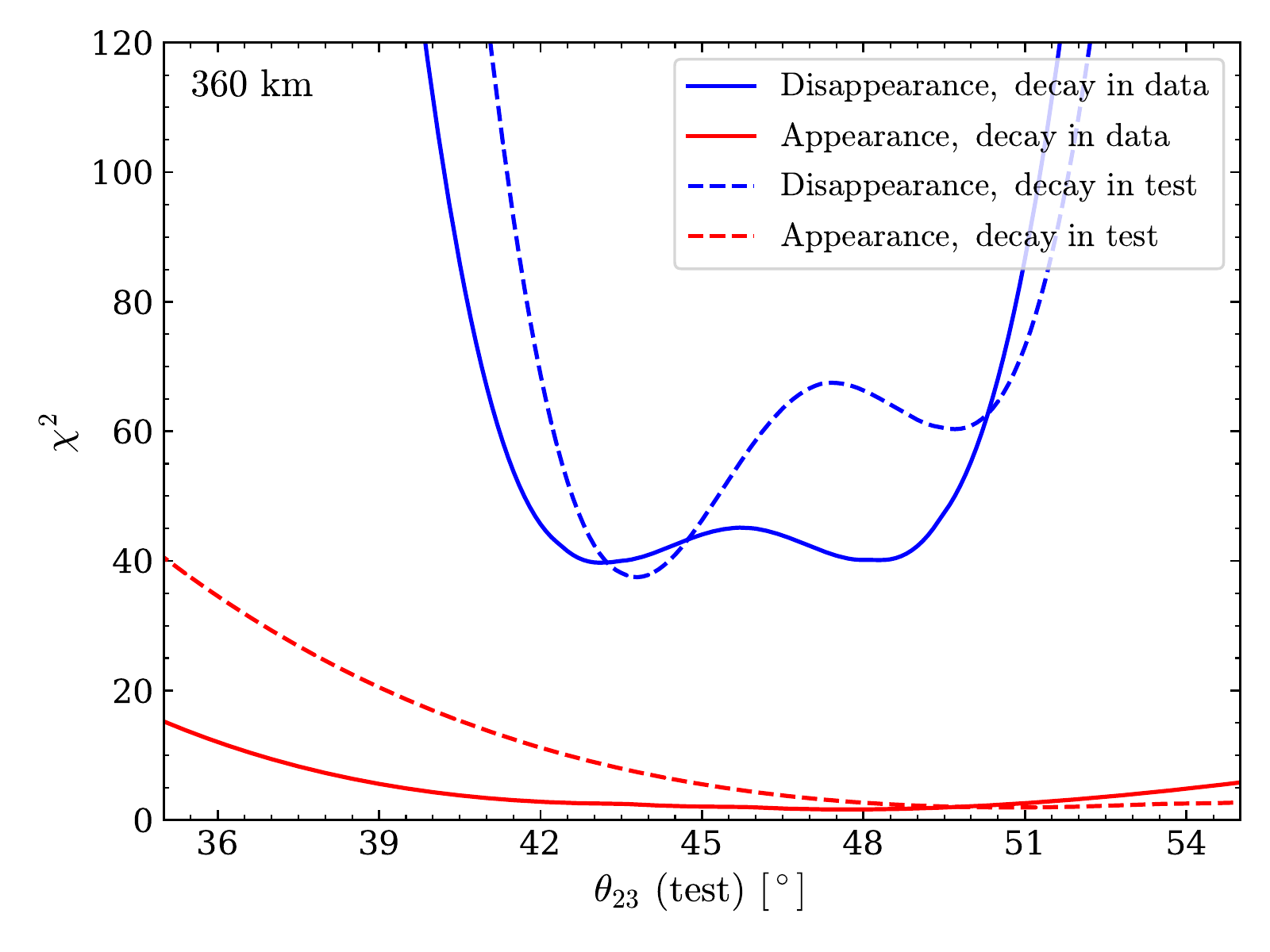}
\end{center}
\vspace{-5mm}
\caption{Contribution to $\chi^2$ from disappearance (appearance) channel displayed in blue (red) at different values of $\theta_{23}$ in test. The case with decay only in data (test) is plotted with a solid (dashed) curve. The value of the decay parameter is $\tau_3/m_3 = 1.0 \times 10^{-11}$~s/eV. Results for $L = 540$~km ($L = 360$~km) are shown in the left (right) panel. The $\chi^2$ is calculated for appearance channel and disappearance channels separately for different values of $\theta_{23}$ in the test, while all other parameters are kept fixed.}
\label{fig_chitheta1}
\end{figure}

\begin{figure}
\begin{center}
\includegraphics[width=0.49\textwidth]{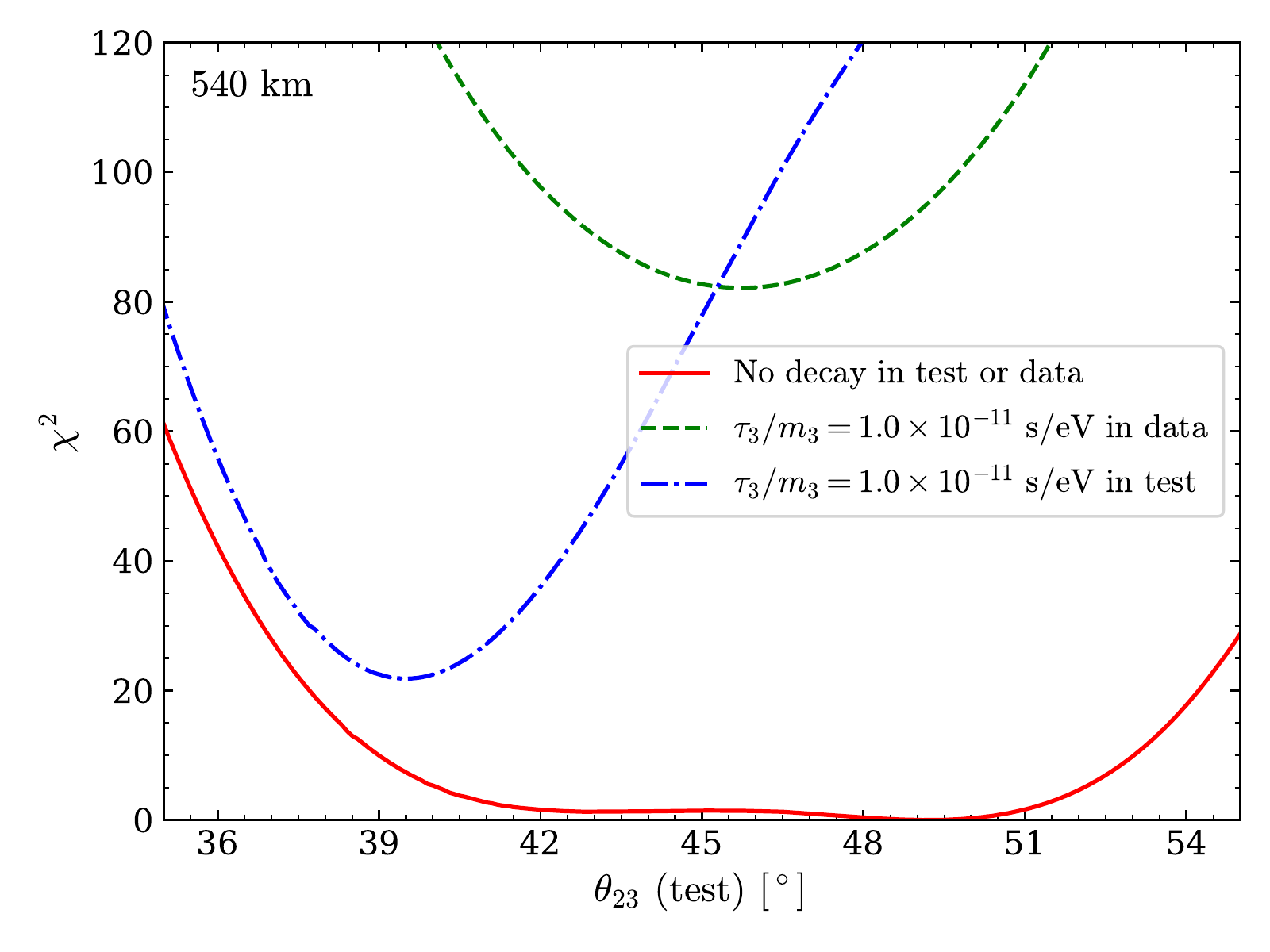} 
\includegraphics[width=0.49\textwidth]{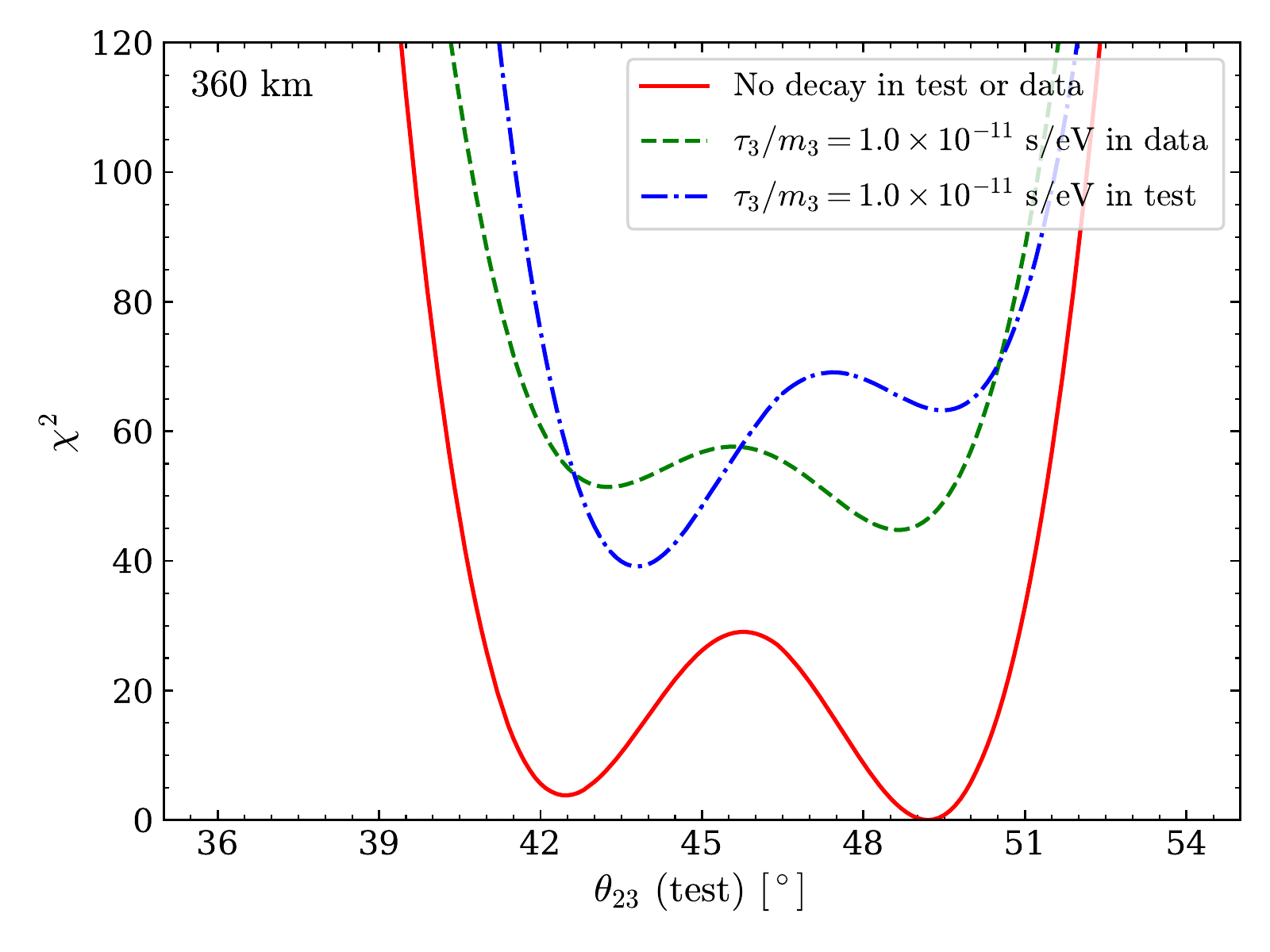} 
\end{center}
\vspace{-5mm}
\caption{Effect on $\chi^2$ from varying $\theta_{23}$ in test. The red solid curve shows the standard case with no decay, the green dashed curve shows the case with decay only in data, and the blue dot-dashed curve shows the case with decay only in test. The value of the decay parameter is $\tau_3/m_3 = 1.0 \times 10^{-11}$~s/eV. Results for $L = 540$~km ($L = 360$~km) are shown in the left (right) panel. The $\chi^2$ is calculated for different values of $\theta_{23}$ in the test, while all other parameters are kept fixed.}
\label{fig_chitheta2}
\end{figure}

In Fig.~\ref{fig_chitheta1}, we study the impact of $\theta_{23}$ marginalization further and show the individual $\chi^2$ for the appearance (red curves) and disappearance (blue curves) channels as functions of $\theta_{23}$ assumed in the test. The dashed curves are for the case of discovery, whereas the solid curves are relevant for sensitivity. The left panel is displayed for the 540~km baseline option, while the right panel for the 360~km one. The data for all cases are generated at $49.2^\circ$. The effect of the appearance channel is negligible, as was pointed out earlier, so we concentrate only on the disappearance channel. We note that for the 540~km baseline option, the best fit for the sensitivity case comes in the lower octant, while the value of $\theta_{23}=49.2^\circ$ is strongly disfavored. For the discovery case, the effect is less dependent on $\theta_{23}$, but the effect of marginalization exists. This is consistent with our discussion in the previous paragraphs in connection with $\Delta P_2$ in Fig.~\ref{fig_terms}. 

In Fig.~\ref{fig_chitheta2}, we show the combined $\chi^2$ coming from both disappearance and appearance channels as a function of $\theta_{23}$ in the test. The left panel is displayed for the 540~km baseline option, while the right panel for the 360~km one. The data are generated for $\theta_{23}=49.2^\circ$. We show this for three different cases -- the red solid curves show the $\chi^2$ for the case when we have standard oscillations in both data and theory. Therefore, these curves show the octant sensitivity for the 360~km and 540~km baseline options. We observe that the octant sensitivity of the 360~km baseline option is better than that of the 540~km one. Indeed, the octant sensitivity of the 540~km baseline option is rather poor. The blue dot-dashed curves show the $\chi^2$ for the case when we have no decay in data and decay in theory. This is the case for the sensitivity study.  The low $\theta_{23}$ octant sensitivity of the 540~km baseline option results in the true octant getting disfavored at a very high significance level in the sensitivity case. With decay switched on in the theory, the fit prefers to choose the wrong octant to lower the $\chi^2$ coming from the inclusion of the decay parameter in the fit. The reason for choosing lower values of $\theta_{23}$ in the fit for the sensitivity case has been discussed in the previous paragraphs. For the 360~km baseline option, the fit also prefers the wrong $\theta_{23}$ octant for the blue dot-dashed curve, but the $\chi^2$ difference between the two octants is lower than for the 540 km case. 

Let us now briefly compare the bound on the decay parameter expected for the ESSnuSB experiment, with the corresponding values for other accelerator, atmospheric, and reactor neutrino experiments. In Table~\ref{Tab:bound}, we list the (expected) bounds on $\tau_3/m_3$ from different experiments along with what has been obtained in this work.
\begin{table}
	\centering
	\setlength{\extrarowheight}{0.1cm}
	\begin{tabular}{|c|c|c|}
		\hline
		 Experiment & $90~\%$ C.L.~($3 \sigma$) bound on $\tau_3/m_3$ [s/eV] & Ref.\\
		\hline
		T2K + NO$\nu$A &  $2.3~(1.5)  \times 10^{-12}$ & \cite{Choubey:2018cfz}\\
		T2K + MINOS &  $2.8~(1.8)  \times 10^{-12}$ & \cite{Gomes:2014yua}\\
		SK + MINOS & $2.9~(0.54)  \times 10^{-10}$ & \cite{GonzalezGarcia:2008ru}\\
		\hline
		MOMENT &  $2.8~(1.6)  \times 10^{-11}$ & \cite{Tang:2018rer}\\
		ESSnuSB (540~km) & $4.22~(1.68)  \times 10^{-11}$ & This work\\
		DUNE (CC) &  $4.50~(2.38)  \times 10^{-11}$ & \cite{Choubey:2017dyu}\\
		ESSnuSB (360~km) &  $4.95~(2.64)   \times 10^{-11}$ & This work \\
		 DUNE (CC + NC) & $5.1~(2.7)  \times 10^{-11}$ & \cite{Ghoshal:2020hyo}\\
		JUNO &  $9.3~(4.7)  \times 10^{-11}$ & \cite{Abrahao:2015rba} \\
		INO &  $1.51~(0.566)   \times 10^{-10}$ & \cite{Choubey:2017eyg}\\
		KM3NeT-ORCA &  $2.5~(1.4)  \times 10^{-10}$ & \cite{deSalas:2018kri} \\
		\hline
	\end{tabular}
	\caption{Comparison of bounds on $\tau_3/m_3$ from different experiments. The bounds for T2K + NO$\nu$A, T2K + MINOS, and SK+MINOS are obtained using real data. The abbreviation CC stands for an analysis with charged current events and the abbreviation CC+NC stands for an analysis with combined charged current and neutral current events.}
	\label{Tab:bound}
\end{table}

\begin{table}[H]
\hspace{-1.2 cm}
	\setlength{\extrarowheight}{0.1cm}
	\begin{tabular}{|p{0.18\textwidth}|p{0.18\textwidth}|p{0.05\textwidth}|p{0.05\textwidth}|p{0.04\textwidth}|p{0.05\textwidth}|p{0.06\textwidth}|p{0.06\textwidth}|p{0.19\textwidth}|}
		\hline
		 Experiment & Specifications & \multicolumn{6}{c|}{Values of true parameters} & Parameters minimized in the fit \\
		 \cline{3-8}
		  &  & $\theta_{12}$ & $\theta_{13}$ & $\theta_{23}$ & $\Delta m^2_{21}$ & $\Delta m^2_{31}$ & $\delta_{\rm CP}$  & \\
		  &  & $[^\circ]$ & $[^\circ]$ & $[^\circ]$ & [$10^{-5}$ eV$^2$] & [$10^{-3}$ eV$^2$] & $[^\circ]$  & \\
		\hline
		T2K + NO$\nu$A &  T2K data as in \cite{Abe:2017vif}, NO$\nu$A data as in \cite{Adamson:2017gxd} & $~~~$- & $~~$- & $~~$- & $~~$- & $~~~$-  & $~~~$- & $\theta_{23}$,  $\delta_{\rm CP}$, $\Delta m^2_{31}$ \\
		T2K + MINOS &  T2K disappearance data as in \cite{Abe:2014ugx}, MINOS data as in \cite{Adamson:2010wi,Adamson:2013whj} & $~~~$- & $~~$- & $~~$- & $~~$- & $~~~$-  & $~~~$- & $\theta_{23}$, $\Delta m^2_{31}$ \\
		SK + MINOS &  SK data as in \cite{Fukuda:1998mi,Ashie:2005ik,Hosaka:2006zd}, MINOS data as in \cite{nelson_talk,weber_talk} & $~~~$- & $~~$- & $~~$- & $~~$- & $~~~$-  & $~~~$- & $\theta_{23}$, $\Delta m^2_{31}$ \\
		\hline
		MOMENT & Gd-doped 500 kt water-Cherenkov, 5 yr ($\nu$)+ 5 yr ($\bar{\nu}$) & $33.82$ & $8.61$ & $49.6$ & $7.39$ & $2.525$  &  $-90$ & all \\
		ESSnuSB & 507 kt water-Cherenkov, 5 yr ($\nu$)+ 5 yr ($\bar{\nu}$) & $34.44$ & $8.57$ & $49.2$ & $7.42$ & $2.517 $ &  $-163$  &  $\theta_{13}$, $\theta_{23}$, $\delta_{\rm CP}$\\
		DUNE (CC) &  40 kt liquid-Argon time projection chamber, 5 yr ($\nu$)+ 5 yr ($\bar{\nu}$) & $34.8$ & $8.5$ & $42$ & $7.5$ & $2.457$ &  $-90$ & $\theta_{13}$, $\theta_{23}$, $\delta_{\rm CP}$, $\Delta m^2_{31}$ \\
		DUNE (CC+NC) &  40 kt liquid-Argon time projection chamber, 3.5 yr ($\nu$)+ 3.5 yr ($\bar{\nu}$) & $33.82$ & $8.61$ & $48.3$ & $7.39$ & $2.523$ &  $-138$ & all \\
		JUNO &  20 kt liquid scintillator, 5 yr & $33.5$ & $8.491$ & $~~$- & $7.5$ & $2.46$ &  $~~~$- & all \\
		INO &  50 kt iron-calorimeter, 10 yr & $34.5$ & $8.5$ & $45.0$ & $7.6$ & $2.366$ &  $0$ & $\theta_{13}$, $\theta_{23}$, $\Delta m^2_{31}$\\
		KM3NeT-ORCA & 6 Mt, 10 yr & $34$ & $8.451$ & $47.7$ & $7.55$ & $2.50$ &  $-122$ & $\theta_{23}$, $\Delta m^2_{31}$  \\
		\hline
	\end{tabular}
	\caption{Details of the input parameters, which are used to estimate the bounds on the decay parameter $\tau_3/m_3$.}
	\label{input}
	\end{table}

In this table, the bounds for T2K + NO$\nu$A, T2K + MINOS, and SK+MINOS are obtained using real data, whereas for the other experiments, the bounds correspond to numerical simulations obtained by different groups. 
To understand under which assumptions the bounds are obtained, we list the input parameters for different experiments in Table~\ref{input}. For the bounds, which were obtained from real data, we provide the references of the data for the relevant experiments as well as the parameters that are minimized during the fits. For the bounds, which were estimated using simulated data, we provide the details of the fiducial volume, the run-time, and the true values of the oscillation parameters as well as the parameters that are minimized in the fits for the relevant experiments. As the bounds for T2K + NO$\nu$A, T2K + MINOS, and SK+MINOS are obtained using real data and all relevant parameters are minimized in the fits, these bounds are robust. For MOMENT, ESSnuSB, DUNE, JUNO, INO, and KM3NeT-ORCA, the bounds mainly depend on the true values of the oscillation parameters, the fiducial volume, and the run-time. For different values of these parameters, the bounds can be different. In particular, we note that except for $\theta_{23}$ and $\delta_{\rm CP}$, the true values of the oscillation parameters are very similar for all experiments. Since the sensitivity mainly comes from the disappearance channel, the bounds are not expected to change much with respect to different values of $\delta_{\rm CP}$. However, different true values of $\theta_{23}$ can alter the sensitivity. Furthermore, depending on the implementation of systematics, backgrounds, and efficiencies, the bounds can also change. For example, the bounds of T2K + NO$\nu$A, MOMENT, ESSnuSB, and DUNE are estimated using the GLoBES software, whereas for T2K + MINOS, SK+MINOS, JUNO, INO and KM3NeT-ORCA, the bounds are estimated using other software. Nevertheless, we expect the order of magnitude to be same even if the values of the input parameters or the implementation of the detector response are changed.

From Table~\ref{Tab:bound}, we note that the bounds obtained from the ongoing atmospheric neutrino experiment SK along with MINOS, future atmospheric experiment INO, and the atmospheric data of the future ultra-high energy neutrino experiment KM3NeT-ORCA are one order of magnitude stronger than the bounds obtained from future accelerator and reactor experiments. On the other hand, the bound obtained from the currently running accelerator experiments T2K and NO$\nu$A, along with MINOS, are one order of magnitude lower than the expected bounds from future accelerator and reactor experiments. Among ESSnuSB, DUNE, MOMENT, and JUNO, the expected bound for the JUNO experiment is the strongest and the one for MOMENT is weakest. The sensitivity of ESSnuSB is slightly better than that of DUNE obtained from a charged current analysis, for the baseline option of 360~km and worse for the baseline option of 540~km, but better than the one of MOMENT. Note that a stronger bound for DUNE could possibly be obtained in a combined analysis of both charged current and neutral current events.

\subsection{Precision measurement of the decay parameter}

In this section, we discuss the capability of ESSnuSB to measure the value of the decay parameter $\tau_3/m_3$, if there exists invisible decay in Nature. In Fig.~\ref{fig_prec}, we plot the precision $\chi^2$ as a function of $\tau_3/m_3$ (test) for three different true values of $\tau_3/m_3$.
\begin{figure}
\begin{center}
\includegraphics[width=0.49\textwidth]{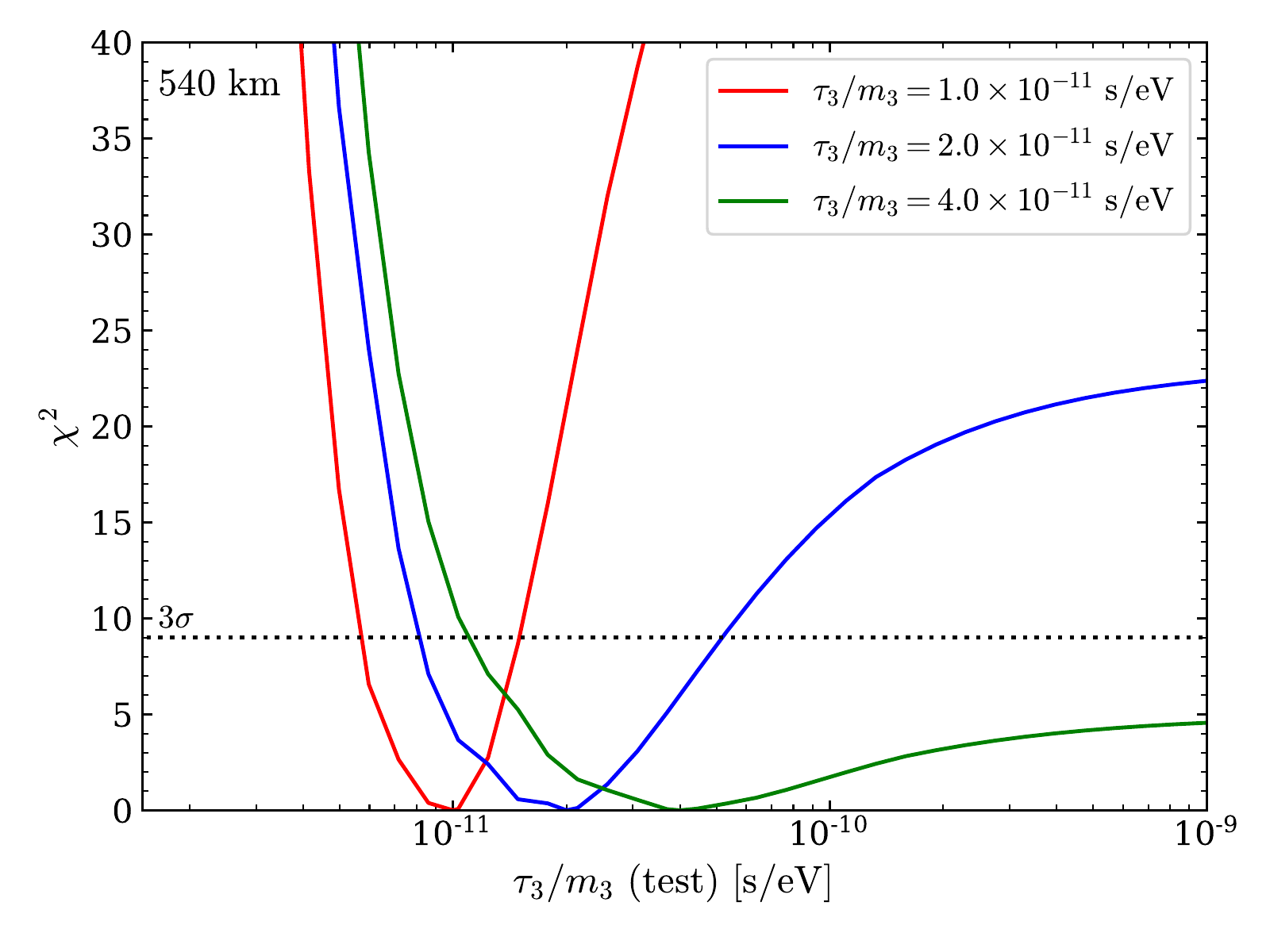} 
\includegraphics[width=0.49\textwidth]{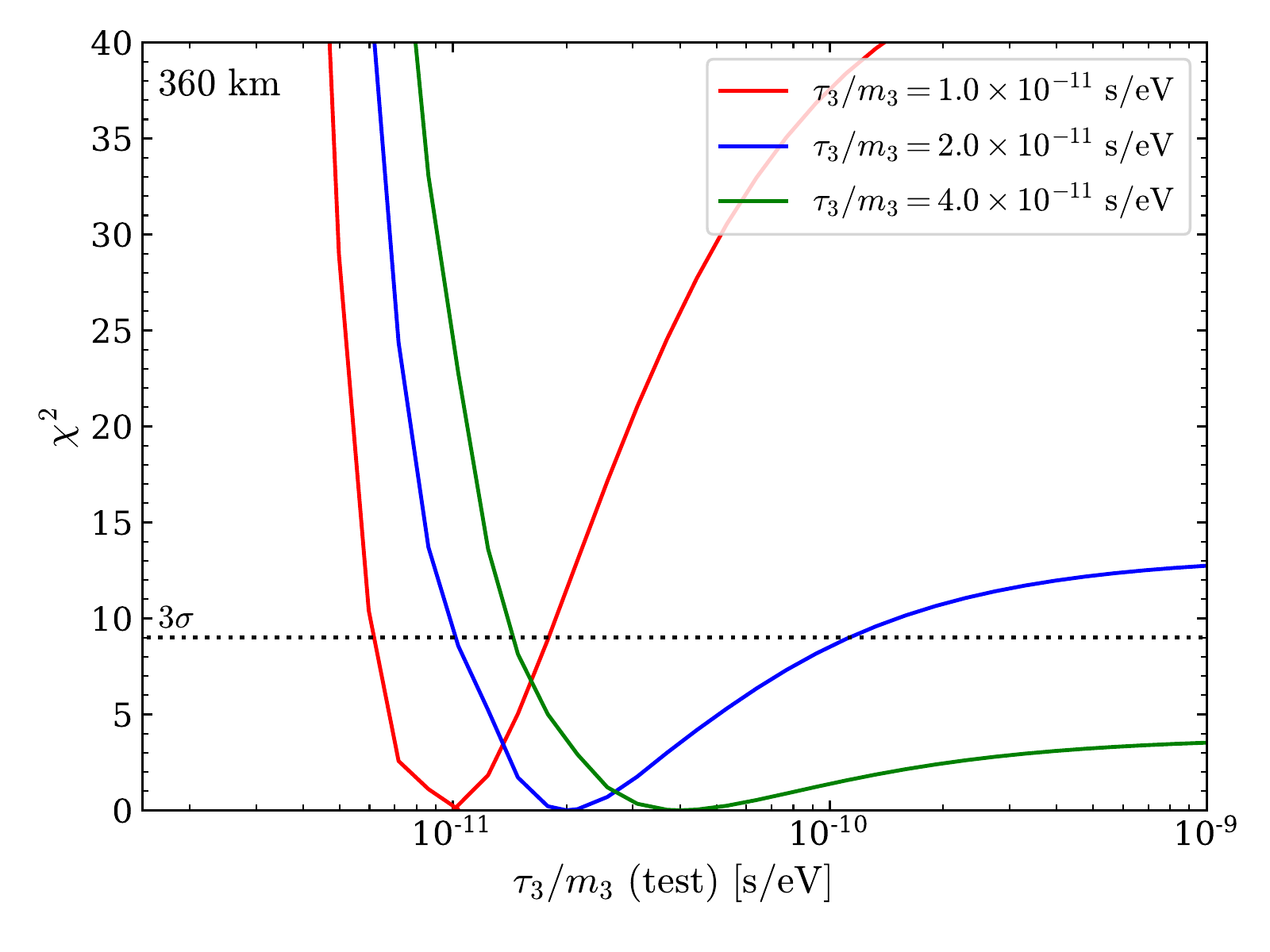} 
\end{center}
\vspace{-5mm}
\caption{Precision $\chi^2$ as a function of $\tau_3/m_3$ (test) for three different values of $\tau_3/m_3$ (true).}
\label{fig_prec}
\end{figure}
The left panel is for $L = 540$~km and the right panel is for $L = 360$~km. In each panel, the red, blue, and green curves correspond to the true value of $\tau_3/m_3$ of $1.0 \times 10^{-11}$~s/eV, $2.0 \times 10^{-11}$~s/eV, and $4.0 \times 10^{-11}$~s/eV, respectively. We choose the values of $\tau_3/m_3$ such that ESSnuSB is sensitive to these values and can distinguish decay and the standard neutrino oscillations within $3 \sigma$. We note that as the value of the decay parameter increases, ESSnuSB becomes comparatively less sensitive to decay, and therefore, the precision decreases. Between $L = 540$~km and $L = 360$~km, the precision capability is better for the 540~km baseline option of ESSnuSB. For $L = 540$~km ($L = 360$~km), the allowed ranges of $\tau_3/m_3$ at $3 \sigma$ are 
\begin{align}
& [5.63 \times 10^{-12}, 1.59 \times 10^{-11}]~\mbox{s/eV} \qquad ([6.10 \times 10^{-12}, 1.80 \times 10^{-11}]~\mbox{s/eV}), \nonumber\\
& [8.06 \times 10^{-12}, 5.18 \times 10^{-11}]~\mbox{s/eV} \qquad ([1.01 \times 10^{-11}, 1.13 \times 10^{-10}]~\mbox{s/eV}) \nonumber
\end{align}
for the true value of $\tau_3/m_3$ of $1.0 \times 10^{-11}$~s/eV and $2.0 \times 10^{-11}$~s/eV, respectively.

\subsection{CP sensitivity in presence invisible neutrino decay}
\label{CP}

In this section, we discuss the CP violation discovery and the CP precision sensitivity of ESSnuSB in presence of invisible neutrino decay. The CP violation discovery potential of an experiment is defined as its capability to distinguish a true value of $\delta_{\rm CP}$ from $0^\circ$ and $180^\circ$, whereas the CP precision capability of an experiment is defined as its capability to exclude all values of $\delta_{\rm CP}$  other than the true value. In the upper panels of Fig.~\ref{fig_cp}, we present the CP violation discovery $\Delta \chi^2$ as a function of the true $\delta_{\rm CP}$, and in the lower panels, we present the $1 \sigma$ precision as a function of the true $\delta_{\rm CP}$. In each row, the left panel is for $L = 540$~km and the right panel is for $L = 360$~km. In all panels, the red curve corresponds to the standard neutrino oscillation scenario, where there is no decay in both data and theory. To study the effect of decay in the measurement of $\delta_{\rm CP}$, we consider two scenarios: (i)~the presence of decay in both data and theory (blue curve) and (ii)~the presence of decay in data but not in theory (green curve). Scenario~(i) addresses the question how the CP sensitivity of ESSnuSB will be altered if there exists decay in Nature and scenario~(ii) addresses the question what happens if we try to fit a theory without decay to a data set that includes decay. Scenario~(ii) is usually the case because when a data set from a neutrino oscillation experiment is available, it will first be fitted using the standard neutrino oscillation framework. The value of the decay parameter is $\tau_3/m_3 = 1.0 \times 10^{-11}$~s/eV and to generate the blue curves, we minimize $\tau_3/m_3$ in the $3 \sigma$ range as obtained from Fig.~\ref{fig_prec}. When decay is present in both data and theory, the CP violation sensitivity is better than for the standard scenario around $\delta_{\rm CP} = \pm 90^\circ$. The effect is larger for $L = 540$~km compared to $L = 360$~km. For CP precision, the sensitivity is weaker for the true $\delta_{\rm CP}$ values of $\pm 90^\circ$ and stronger for the true $\delta_{\rm CP}$ values of $0^\circ$ and $180^\circ$ as compared to the CP precision in the standard three-flavor neutrino oscillation scenario. This deviation is larger for $L = 360$~km than for $L = 540$~km. Further, the CP violation sensitivity due to standard neutrino oscillations and the sensitivity due to the presence of decay in data but not in theory are almost identical for  $\delta_{\rm CP} = 90^\circ$. However, for $\delta_{\rm CP} = - 90^\circ$, the CP violation sensitivity due to the presence of decay in data is better than for the standard case. These conclusions are true for both baseline length options of ESSnuSB. In this case, the value of $\chi^2_{\rm min}$ is 45.4 for $L = 360$~km and 82.2 for $L = 540$~km. Nevertheless, when decay is present in theory but not in data, the sensitivity to precision is significantly different compared to the standard scenario. This can be qualitatively understood in the following way. When there is decay or no decay in both data and theory, the $\chi^2$ minimum for CP precision is always zero and it appears with the true value of $\delta_{\rm CP}$. However, if there is decay in data and not in theory, the $\chi^2$ minimum is non-zero and it can appear with a different value of $\delta_{\rm CP}$ other than the true value of $\delta_{\rm CP}$ and this can change the sensitivity significantly. This can be viewed as a signature of the presence of decay that, if there exists decay in data and one tries to fit a theory without decay, the CP precision will be significantly different.
\begin{figure}
\begin{center}
\includegraphics[width=0.49\textwidth]{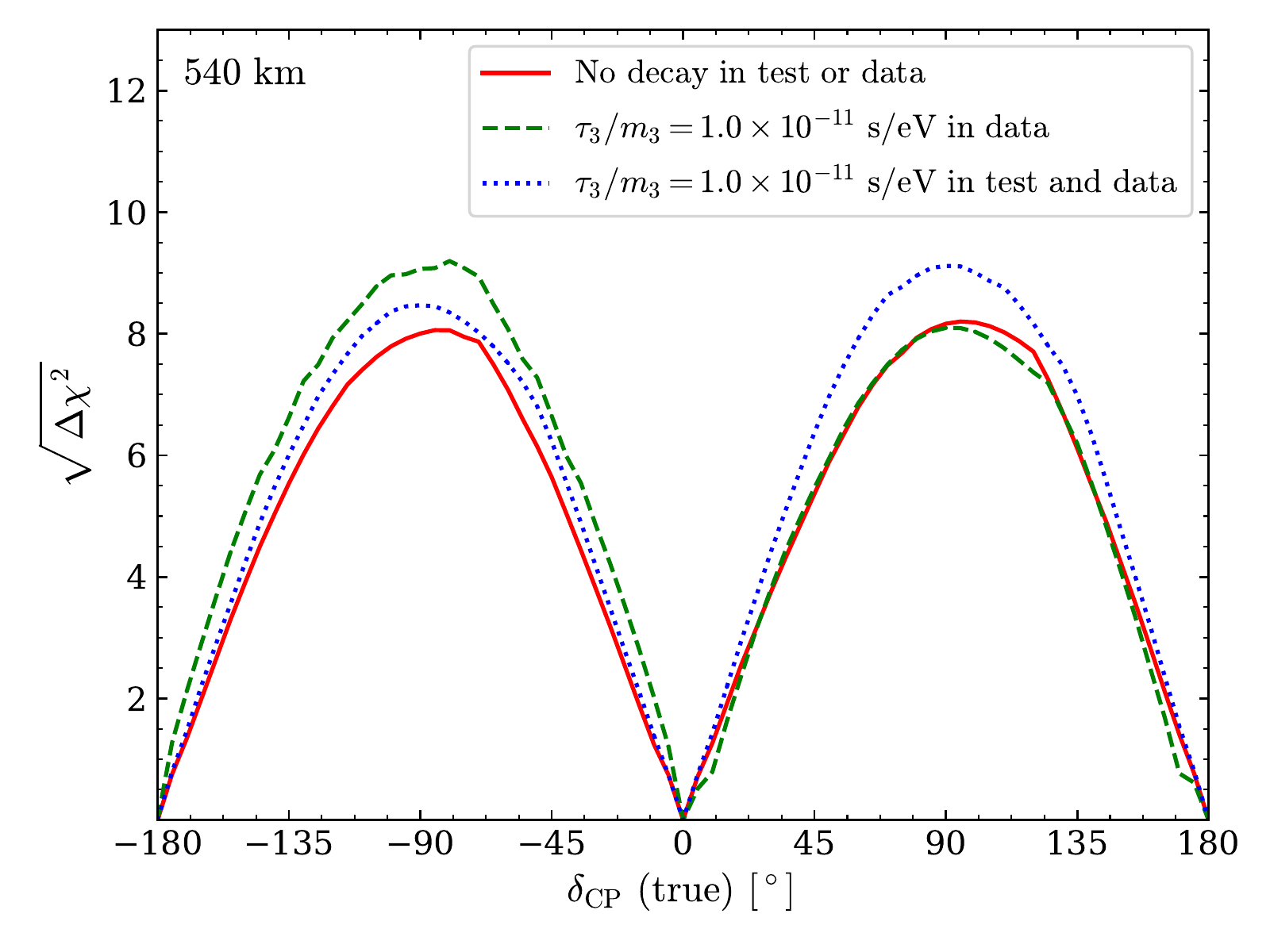} 
\includegraphics[width=0.49\textwidth]{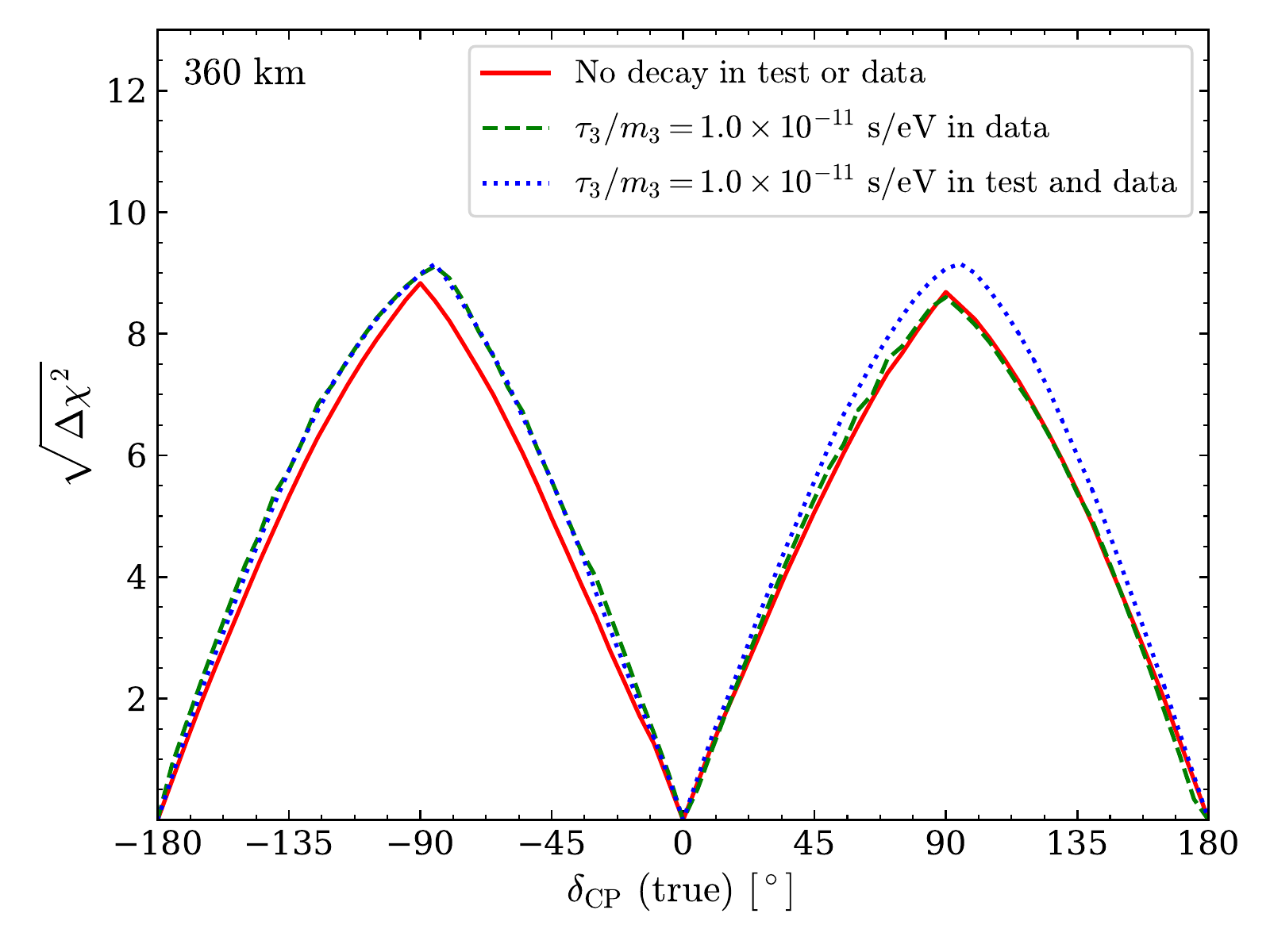} 
\includegraphics[width=0.49\textwidth]{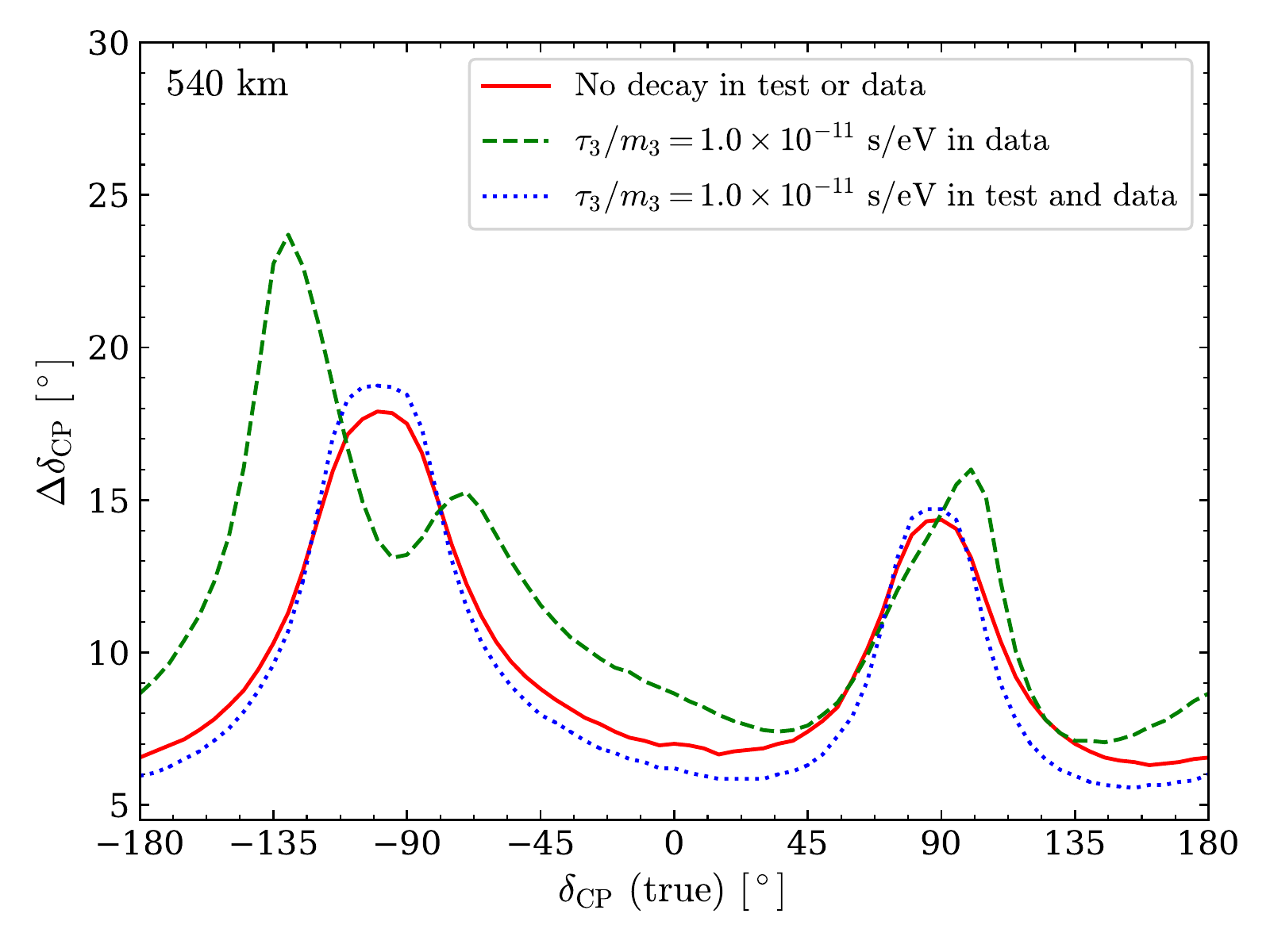} 
\includegraphics[width=0.49\textwidth]{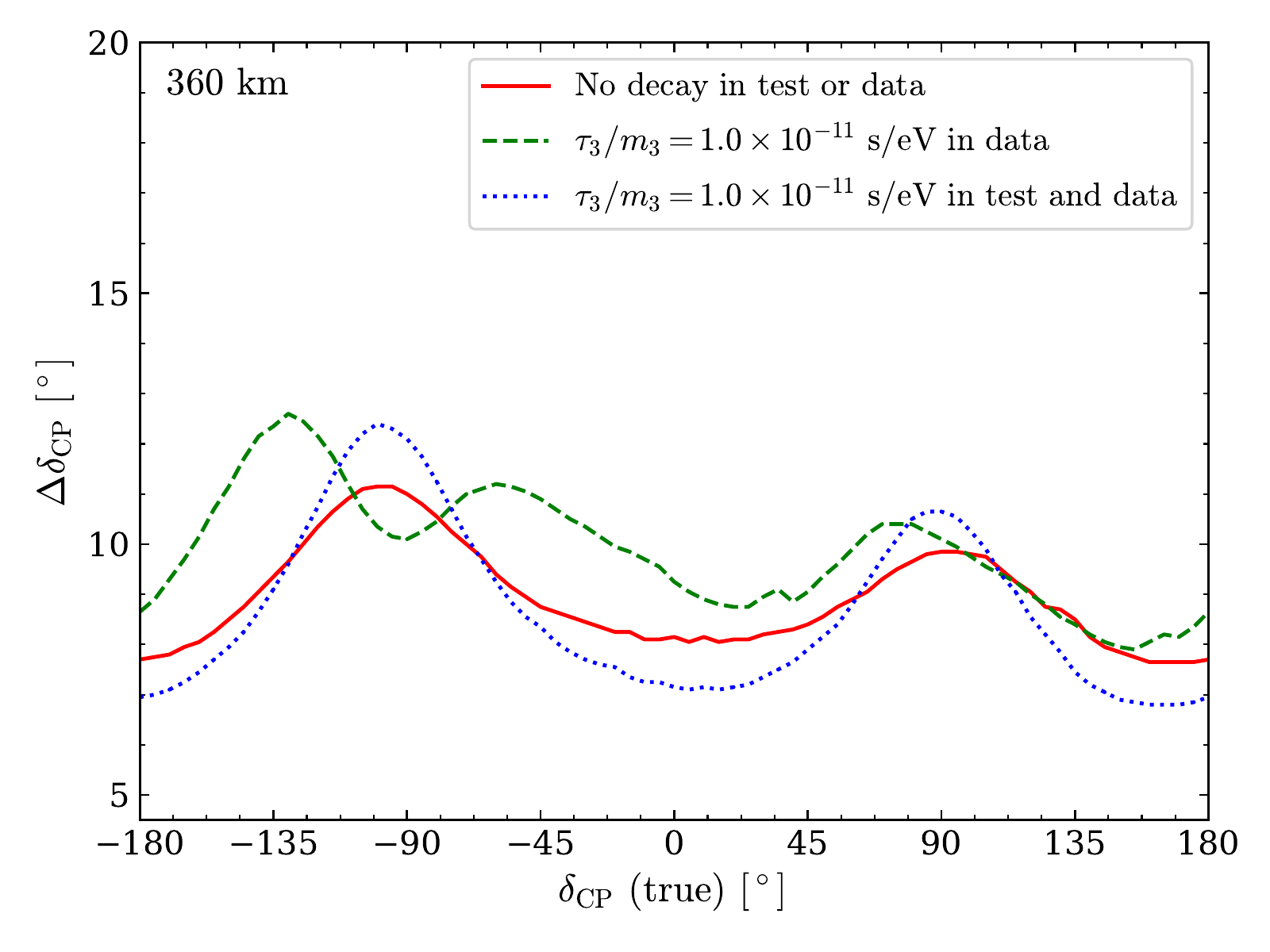} 
\end{center}
\vspace{-5mm}
\caption{CP violation $\chi^2$ as a function of $\delta_{\rm CP}$ true (upper panels) and CP precision $\chi^2$ as a function of $\delta_{\rm CP}$ true (lower panels).}
\label{fig_cp}
\end{figure}

\section{Summary and conclusions}
\label{sec5}

Invisible decay is defined as the decay of a heavy neutrino state into a lighter sterile neutrino state. Due to the presence of invisible neutrino decay, the evolution equation of neutrino propagation is altered, and therefore, it is possible to probe invisible decay in neutrino oscillation experiments. The bound on the decay parameters due to decay of $\nu_1$ and $\nu_2$ comes from the solar and supernova neutrino experiments and the decay due to $\nu_3$ can be probed in current and future accelerator, atmospheric, ultra-high energy, and reactor neutrino experiments. In this work, we have studied the physics sensitivity of the ESSnuSB experimental setup in presence of invisible decay. The primary aim of the ESSnuSB experiment is to measure the leptonic CP-violating phase $\delta_{\rm CP}$ at the second oscillation maximum. In the current work, we have considered two baseline options of ESSnuSB, which are 540~km and 360~km, respectively, and studied (i)~the capability to put bounds on the decay para\-meter, (ii)~the capability to discover invisible decay, (iii)~the bounds on the decay parameter obtained with other experiments, (iv)~the potential to measure a particular value of decay parameter, and (v)~the effect of decay on the measurement of $\delta_{\rm CP}$. In our work, we have shown that the capability of ESSnuSB to put bounds on the decay parameter or the sensitivity $\chi^2$ is better for the baseline option of 360~km, while its capability to discover decay with a particular value of decay parameter or the discovery $\chi^2$ is better for the baseline option of 540~km. The sensitivity $\chi^2$ is worse for the baseline option of 540~km due to the effect of $\theta_{23}$ on decay. Our results show that the bound obtained for ESSnuSB with the baseline option of 360~km is better than the one of DUNE~(CC) and the bound obtained with the baseline option of 540~km is worse than the one of DUNE~(CC), but better than the one of MOMENT. Note that the robustness of the bounds obtained from different experiments depends on experimental specifications, true value of the oscillation parameters, etc. Therefore, the conclusion that has been derived in this work can change based on different assumptions upon which the bounds have been estimated. Assuming true values of the decay parameter outside the bounds obtained for ESSnuSB, we have shown that the capability of ESSnuSB to precisely measure a value of the decay parameter is better for $L = 540$~km than for $L = 340$~km and as the lifetime of decay increases, precision becomes worse. Regarding the measurement of $\delta_{\rm CP}$, we have found that in presence of decay in both data and theory, the CP violation sensitivity is better for $\delta_{\rm CP} = \pm 90^\circ$, whereas the CP precision capability is stronger for $\delta_{\rm CP} = 0^\circ$ and $180^\circ$ and weaker for $\delta_{\rm CP} = \pm 90^\circ$ compared to the standard three-flavor neutrino oscillation scenario. For CP violation sensitivity, the deviation is larger for $L = 540$~km, whereas for CP precision capability, the deviation is larger for $L = 360$~km. Further, if we try to fit data, which include invisible neutrino decay, with a theory that does not incorporate neutrino decay, then there exist a visible difference in the CP violation sensitivity due to effect of decay around $\delta_{\rm CP} = -90^\circ$, whereas the CP precision capability is largely affected as compared to the standard case. The significant change in the CP precision capability can be viewed as a signature of the presence of decay when one tries to fit data that contain decay with a theory that does not include decay. In summary, our results have shown that ESSnuSB provides a good opportunity to study invisible neutrino decay and the primary goal of the ESSnuSB to measure $\delta_{\rm CP}$ can be affected due to the presence of such decay.

\section*{Acknowledgements}

This project is supported by the European Union's Horizon 2020 research and innovation programme under grant agreement No.~777419. T.O.~acknowledges support by the Swedish Research Council (Vetenskapsr{\aa}det) through Contract No.~2017-03934 and the KTH Royal Institute of Technology for a sabbatical period at the University of Iceland. 

\bibliographystyle{JHEP}
\bibliography{ess_decay}
  
\end{document}